\documentclass[useAMS,usenatbib]{mn2e}

\usepackage{graphicx}
\usepackage{times}
 \usepackage{amsmath,amssymb}
\usepackage{epstopdf}

\newif\ifAMStwofonts
\AMStwofontstrue

\newcommand{\be}{\begin{equation}}
\newcommand{\ee}{\end{equation}}
\newcommand{\ba}{\begin{eqnarray}}
\newcommand{\ea}{\end{eqnarray}}
\newcommand{\brr}{\begin{array}}
\newcommand{\err}{\end{array}}
\newcommand{\bc}{\begin{center}}
\newcommand{\ec}{\end{center}}

  \newcommand{\iMpc}{\mbox{ Mpc$^{-1}$}}
   \newcommand{\eV}{\mbox{ eV}}

\newcommand{\mincir}{\raise
  -2.truept\hbox{\rlap{\hbox{$\sim$}}\raise5.truept \hbox{$<$}\ }}
\newcommand{\magcir}{\raise
  -2.truept\hbox{\rlap{\hbox{$\sim$}}\raise5.truept \hbox{$>$}\ }}
\newcommand{\siml}{\raise
  -2.truept\hbox{\rlap{\hbox{$\sim$}}\raise5.truept \hbox{$<$}\ }}
\newcommand{\simg}{\raise
  -2.truept\hbox{\rlap{\hbox{$\sim$}}\raise5.truept \hbox{$>$}\ }}

\newcommand{\aap}{A\&A}
\newcommand{\apj}{ApJ}

\newcommand{\mnras}{MNRAS}

\newcommand{\prd}{Phys. Rev. D}
\newcommand{\physrep}{Phys. Rep.}

\title[Parameter forecasts for Planck and galaxy
surveys]{Cosmological parameters after WMAP5: forecasts for Planck
and future galaxy surveys} 
\author[L.P.L. Colombo et al.]{
L.P.L. Colombo$^1$, E. Pierpaoli$^1$, \& J.R.~Pritchard
$^2$\thanks{Hubble Fellow}\\~\\ $^1$ University of Southern
California, Los Angeles, CA, 90089-0484\\ $^2$ Harvard-Smithsonian
Center for Astrophysics, 60 Garden St., Cambridge, MA 02138\\}

\begin{document}

\date{Accepted ???. Received ???; in original form ???}

\maketitle

\begin{abstract}
With its increased sensitivity and resolution, the Planck satellite is
expected to improve the measurement of most cosmological parameters by
several factors with respect to current WMAP results.  The actual
performance however, may depend upon various aspects of the data
analysis. In this paper we analyse the impact of specifics of the data
analysis on the actual final results.  We also explore the synergies
in combining Planck results with future galaxy surveys.  We find that
Planck will improve constraints on most cosmological parameters by a
factor 3--4 and on the tensor--to--scalar ratio $r$ by a factor
9. Also inflationary parameters, like $r$, $n_{\rm s}$ and $n_{run}$, are
practically not degenerate any longer. The tensor spectral index,
however, is little constrained.  A combination of the 70 to 143 GHz
channels will contain about 90\% of all possible information, with 143
GHz polarisation information carrying about half of the constraining
power on $r$.  Also, the error on $r$ degrades by a
factor 2 if no B modes are considered in the analysis.  High--$l$
temperature information is essential for determination of $n_{\rm s}$ and
$\Omega_{\rm b}$, while improving noise properties increase the $l$--range
where Planck would be cosmic variance limited in polarisation,
implying a significant improvement on the determination of $r$, $\tau$
and $A_s$. However, a sub-percent difference in the FWHM used in the
data analysis with respect to the one in the map will result in a bias
for several parameters.  Finally, Planck will greatly help future
missions like LSST and CIP reach their potentials by providing tight
constraints on parameters like $n_{\rm s}$ and $n_{run}$. Considering Planck
together with these probes will help in breaking degeneracies between
$\Omega_{\rm K}$ and $\Omega_\Lambda$ or $\Omega_{\rm dm}$ and $f_\nu$,
resulting in improvements of several factors in the error associated
to these parameters.

\end{abstract}


\section{Introduction}

Observations of the cosmic microwave background (CMB) have driven an
incredible improvement in our understanding of the Universe in the
last few decades. The discovery of CMB anisotropies by the COBE
satellite triggered the planning of new space missions targeted to the
study of CMB anisotropies: WMAP and Planck.  The WMAP satellite has
already delivered results and is still flying, while the Planck
satellite is scheduled to fly next year.  It is therefore appropriate
at this time to use the experience acquired with the WMAP data
analysis in order to gain information on what can be expected in terms
of parameter determination from Planck.

Given the improved technical performances, Planck carries great
expectations and several authors have proposed a plethora of models
that should be constrained with the new Planck data
\citep[e.g.][]{bond:04,burigana:04,perotto:06,white:06,balbi:07,xia:07,gratton:08,lavacca:08a}. In
general, these estimates aim at an extended set of parameters beyond
the minimal cosmological model, but rely on a number of simplifying
assumptions.  Actual constraints will depend sensitively on the
ability of Planck to clean foregrounds from individual frequency
channels, maximising the cosmological information available.

In this paper, we adopt a different perspective focusing on quite
simple cosmological models and then exploring the required Planck
performance needed for useful constraints. We explore the effects of
foreground cleaning restricting the frequency channels and angular
scales available for cosmology, and the effects of beam
degradation. These have differing effects on individual parameter
constraints and weakened experimental performance can enhance
degeneracies between parameters.  These issues have been partly
addressed in the Planck bluebook \citep{bluebook}. This paper is meant
to provide further and more detailed information than that contained
in the bluebook, especially in light of what has been learnt from
WMAP.  At the same time, we hope that our results are more transparent
than those of full experimental simulations, which fully include
experimental imperfections but are restricted in the range of
parameter space explored.

Specifically, we will first consider a minimal set of parameters and
perform a Monte Carlo Markov Chains (MCMC) estimate of the accuracy
Planck will have in determining them, also showing under which
conditions these results are attained.  Then we will consider a more
extended set of parameters including the possible presence of
gravitational waves, curvature, neutrino--related parameters,
quintessence and we will discuss how the previous results are modified
in the presence of this extended set. While an appropriate treatment
of Planck parameter estimation should also contain a general approach
to reionisation \citep{mortonson:08a,colombo:08}, in the following we
will simply consider a sharp reionisation process parametrised only by
the total optical depth $\tau$. As shown in \citet{colombo:08}, such
assumption should not impact the size of errorbars for most
parameters, which is the main interest of this paper. Finally, we will
discuss the role of Planck as support for the success of galaxy
surveys as cosmological probes. Galaxy surveys are able to constrain a
subset of the cosmological parameter set, typically relying on CMB
data to provide information on the remaining parameters and to
significantly reduce degeneracies. Currently, no space CMB mission is
planned after Planck, and therefore the combination of Planck with
external datasets such as galaxy surveys will be a fundamental tool for
estimating the cosmological parameters in the coming years.

The simplified procedure adopted here will by no means be the one
adopted during the actual data analysis.  For example, the effect of
the precise sky cut applied on power spectrum estimation and
likelihood evaluation \citep[e.g.,][]{hivon:02,lewis:02b,brown:05},
the effect of foreground residual on small and large scale
\citep[e.g.,][]{serra:08,betoule:08} or the effect of lensing on the
polarisation power spectrum are all not investigated here
\citep[e.g.,][]{lewis:05,smith:06}. All these extra complications will
however tend to degrade the level of performance for the Planck
satellite. The results presented here should therefore be considered
as a target or an optimistic limit.

 This paper is organised as follows: in section \ref{sec:metWMAP5} we
 will present the method and apply it to a configuration mimicking the
 WMAP 5 yrs data release; in section \ref{sec:Plonly} we will discuss
 the performances expected for Planck for different frequency
 combinations and sets of parameters; section \ref{sec:support} is
 dedicated to investigating the role of Planck as support for other
 missions and finally section \ref{sec:concl} is dedicated to the
 conclusions.

\section{Method and  WMAP--5yr case} \label{sec:metWMAP5}

In this paper we will perform a MCMC analysis of simplified mock data
to estimate Planck performances in parameter estimation, using the
CosmoMC
package\footnote{http://cosmologist.info/cosmomc/}~\citep{lewis:02a}
The availability of fast codes for evaluation of CMB power spectra,
e.g. PICO\footnote{http://cosmos.astro.uiuc.edu/pico/}~\citep{fendt:07},
allows good convergence of the chains to be obtained in $\sim 1$h on modern
office workstations, thus providing a valid alternative to a Fisher
matrix approach, while avoiding some of the Fisher
matrix shortcomings \citep[see, e.g.,][ for a more in depth
comparison of MCMC and Fisher matrix approaches]{perotto:06}.

Specifically, we will consider idealistic simulations of CMB data,
assuming white isotropic noise and Gaussian beams. Real data analysis
will have to deal with complications like anisotropic and correlated
noise, beam systematics, calibration effects. A complete treatment of
these effects is beyond the scope of this paper and to some extent
would require access to the actual measurements and in--flight data.
In addition, we do not consider residual foreground contribution in
the mock data. Rather we assume that some of the frequencies will be
used to completely clean the remaining channels, which will be the
only ones used for data analysis. In practice, this is the strategy
adopted up to now \citep[e.g.][]{dunkley:08}. We will then show how our
results depend on this choice of channels.

For full sky and noiseless data, the exact likelihood of a
cosmological model, defined by a set of CMB angular power spectra
${\bf C_\ell}^{\rm th}$, given a simulated dataset,
$\bf{\hat{C}_\ell}$, is given by the Wishart distribution
\citep{percival:06,hamimeche:08}.  It reads
\begin{eqnarray}
\label{eq:wishart}
\log {\cal L}( {\bf C_\ell}^{\rm th}| {\bf \hat{C}_\ell} ) =  \\
\nonumber
 -\frac{1}{2} \sum_{\ell} (2\ell +1)\left\{ {\rm Tr}\left[{\bf\hat{C}_\ell} 
( {\bf C_\ell}^{\rm th})^{-1} \right]
- \ln |{\bf \hat{C}_\ell} ( {\bf C_\ell}^{\rm th})^{-1}| -n \right\} 
\end{eqnarray}
where $n$ is the number of different modes ($T, E$ and $B$)
considered. With this choice of normalisation $\log {\cal L}( {\bf
  C_\ell}^{\rm th}| {\bf \hat{C}_\ell} ) = 0$ for $ {\bf C_\ell}^{\rm
  th} = {\bf \hat{C}_\ell}$. In the presence of white noise and for an
instrument with Gaussian beams, the above expression holds provided we
replace ${\bf C_\ell}$ with ${\bf C_\ell + {\cal N}_\ell}B^2_\ell $,
where ${\bf{\cal N}}_\ell$ are the white noise spectra and $B_\ell$ is
the spherical harmonic transform of the instrument's beam. In
addition, when combining different frequencies, we use an inverse
noise weighting.

In the presence of sky cuts or non--uniform noise,
equation~\ref{eq:wishart} is no longer exact. In this work, we
approximate the effect of sky cuts by multiplying the r.h.s. in
equation~\ref{eq:wishart} by a factor $f_{\rm sky}^2$, where $f_{\rm
  sky}$ is the fraction of sky actually observed. This {\it ad--hoc}
correction crudely accounts for the loss of statistics and the induced
correlations between different multipoles, and is already implemented
in CosmoMC. For Planck, we assume $f_{\rm sky} = 0.80$.  This
approximation is not well suited to describe the behaviour of the
likelihood function at low multipoles, which are critical for the
determination of cosmological parameters like the optical depth to
reionisation, $\tau$, and the tensor--to--scalar ratio, $r$.  Features
like non--uniform sky coverage alter the shape of the likelihood
function, and introduce correlations between multipoles, in particular
at $\ell$'s corresponding to scales similar to the dimensions of the
cuts. A discussion of how this affect our results for $r$ is presented
in section \ref{sec:r}.  

As a first application, we consider the WMAP 5 yrs (WMAP5) case,
introducing in the analysis the frequencies that were actually used in
the real data analysis at different scales.  In particular, we adopted
V and W bands in temperature and Ka,Q and V bands for large scale
polarisation reducing to Q and V only for small scale polarisation.
For WMAP we considered $75\%$ sky coverage, according to the mask
applied by the WMAP team to the actual observed maps, and an average
noise value for each frequency considered computed from such maps. We
do not consider the effect of lensing or marginalisation over the
amplitude of the Sunayev-Zeldovich template.

The results are summarised in table~\ref{tab:wmap}, where we report
the percentage difference in the parameters' standard deviation with
respect to the actual WMAP5 performances \citep{dunkley:08} for a $\Lambda$CDM
cosmology with tensors. For comparison, we also report the expected
performance of the Planck satellite, considering an ideal combination
of the 70, 100 and 143 GHz channels (see below for more details).  We
conclude that this simplified approach in the case of WMAP5 leads to
underestimate of parameters uncertainties by $25-30\%$. Although this
result does not fully validate this procedure for Planck, it is
encouraging that its performances on simulated WMAP5 data are so close
to the actual measured value.

\begin{figure*}
\centerline{
\includegraphics[width=16.cm]{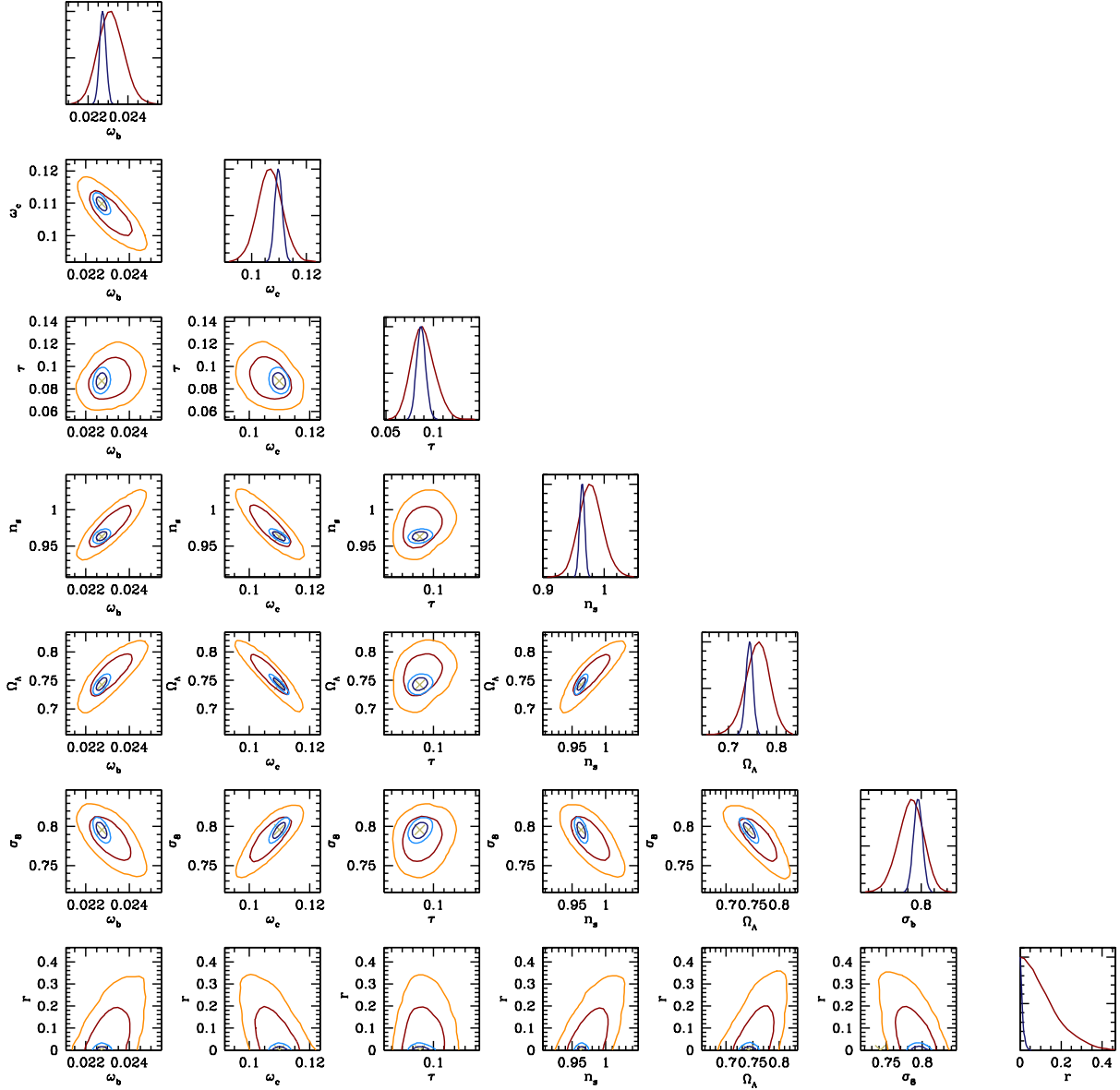}
}
\caption{Marginalised distributions and joint 2D confidence regions
for the base fiducial model analysed assuming a combination of $70,
100$ and $143$ GHz Planck channels (blue/cyan lines) or WMAP 5yr
specifications (red/orange lines).  The 2D plots show the $1$-- and
$2$--$\sigma$ regions; crosses mark the input values for the
parameters.}
\label{fi:2D_base}
\end{figure*}

Figure~\ref{fi:2D_base} provides a visual comparison of our analysis
applied to simulated WMAP5 and Planck data. Planck is expected to
improve on this result by typically a factor 2.5 to 4, except in the
case of $r$, whose upper limit for a fiducial model with $r = 0$
shrinks by almost an order of magnitude. It is also evident that
Planck will remove or strongly reduce several degeneracies which
affect WMAP data, in particular those involving the optical depth to
reionisation, $\tau$. Notice that in the WMAP5 simulations we did not
include the contribution of $B$--modes, while for Planck we also
included simulated BB data. If systematics and/or foregrounds prevent
Planck from measuring the cosmological BB spectrum, the expected
improvement on the tensor--to--scalar ratio upper limit would be by a
factor $\sim 3$ (see section~\ref{sec:r} for a detailed discussion).

\begin{table}
\centerline{
\begin{tabular}{@{}lccc}
\hline
                    & Percentage  & Ratio to     \\
                    & difference  & Planck   \\
\hline
$\omega_{\rm b}$          & 0.22        & 3.9    \\
$\omega_{\rm c}$          & 0.38        & 3.2    \\
$\tau    $          & 0.33        & 2.5    \\
$n_{\rm s}     $          & 0.16        & 4.6    \\
${\cal A}_{\rm s}$        & 0.31        & 2.4    \\
$r$                 & 0.33        & 9.1    \\
$H_0  $             & 0.23        & 3.9    \\
\hline
\end{tabular}
}
\caption{Comparison of uncertainties from actual WMAP 5--year data
  analysis and MCMC estimates. Table shows the percentage difference
  $1 - \sigma_{\rm MCMC} / \sigma_{\rm WMAP}$ for the basic $\Lambda$CDM +
  tensor model. The third column shows the MCMC estimates of the
  standard deviation in unit of the expected Planck accuracy, assuming
  cleaning of the 70, 100 and 143 GHz channels (see
  table~\ref{tab:base70}).}
\label{tab:wmap}
\end{table}

\section{Planck--only constraints}\label{sec:Plonly}

\begin{table}
\centerline{
\begin{tabular}{@{}lccccccc}
\hline
                                 &    & LFI  &  &  &     & HFI&  \\
\hline
Central Frequency                & 30 & 44   & 70& & 100 & 143 & 217   \\
Angular Resolution (arcmin)      & 33 & 24   & 14& & 9.5 & 7.1 & 5.0 \\
$\Delta T / T$ per pixel ($I$)   & 2.0 & 2.7 &4.7& & 2.5 & 2.2 & 4.8 \\
$\Delta T / T$ per pixel ($Q$, $U$)& 2.8 & 3.9&6.7 & & 4.0 & 4.2 & 9.8 \\
\hline
\end{tabular}
}
\caption{Planck specification as reported in the Planck bluebook. 
Listed sensitivities are goal sensitivity  assuming 14 months integration,
for a square pixel with a side equal to the angular resolution
of the corresponding channel.}
\label{tab:bluebook}
\end{table}

Planck has a noise level which is a factor 2-10 lower than WMAP5, a
wider frequency coverage, and a finer angular resolution. Cosmological
constraints for Planck will therefore come from a different set of
scales and frequency ranges than in the case of WMAP.
 
In the following, we will adopt the simplified procedure outlined
above to assess Planck performances under different data analysis
circumstances. We will use the noise specifications derived from the
bluebook, also reported in table~\ref{tab:bluebook}.

Initially we will consider the following fiducial $\Lambda$CDM minimal
model, based on WMAP5 results: flat, $\omega_{\rm b} =0.02273,
\omega_{\rm c} =0.1099, \tau =0.087, n_{\rm s} =0.963, H_0 =71.9$
Km/s/Mpc and no tensor perturbations. The amplitude of the scalar
fluctuation spectrum, $A_s$, derives from the requirement that
$\sigma_8 =0.796$. Following CosmoMC convention, we define here the
tensor--to--scalar ratio $r$ as the ratio of the initial curvature and
gravitational waves power spectra at a reference scale of $k_{\rm piv}
= 0.05 {\rm Mpc}^{-1}$. In the rest of the paper, we define ${\cal
A}_{\rm s} \equiv {\rm Log}(10^{10}A_s)$. In order to explore the
parameter space in a more efficient way, it is useful to chose as a
primary parameter the angle subtended by the sound horizon at
recombination, $\theta$.  Our basic parameter set is then:
\begin{equation}
\left\{\omega_{\rm b}, \omega_{\rm c}, \theta, \tau, n_{\rm s}, {\cal
A}_{\rm s}, r \right\}~.
\end{equation}
Notice that we include $r$ in the analysis, even if the fiducial model
has no tensors (see section \ref{sec:r} for a discussion of models
with non--zero $r$). With this choice of parametrisation, $H_0$ and
$\sigma_8$ are then derived parameters.

\begin{figure*}
\centerline{
\includegraphics[width=6.cm]{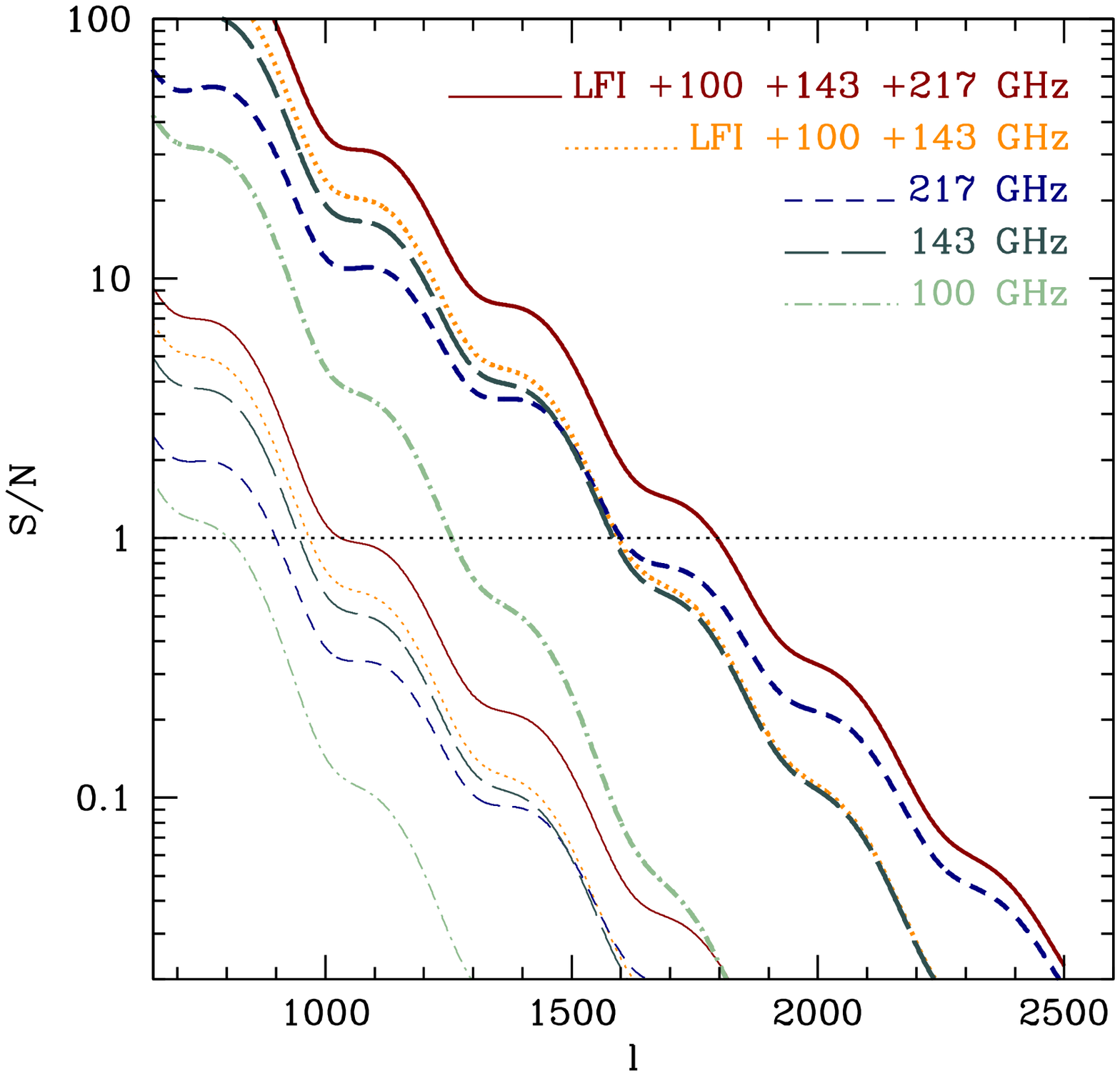}
\includegraphics[width=6.14cm]{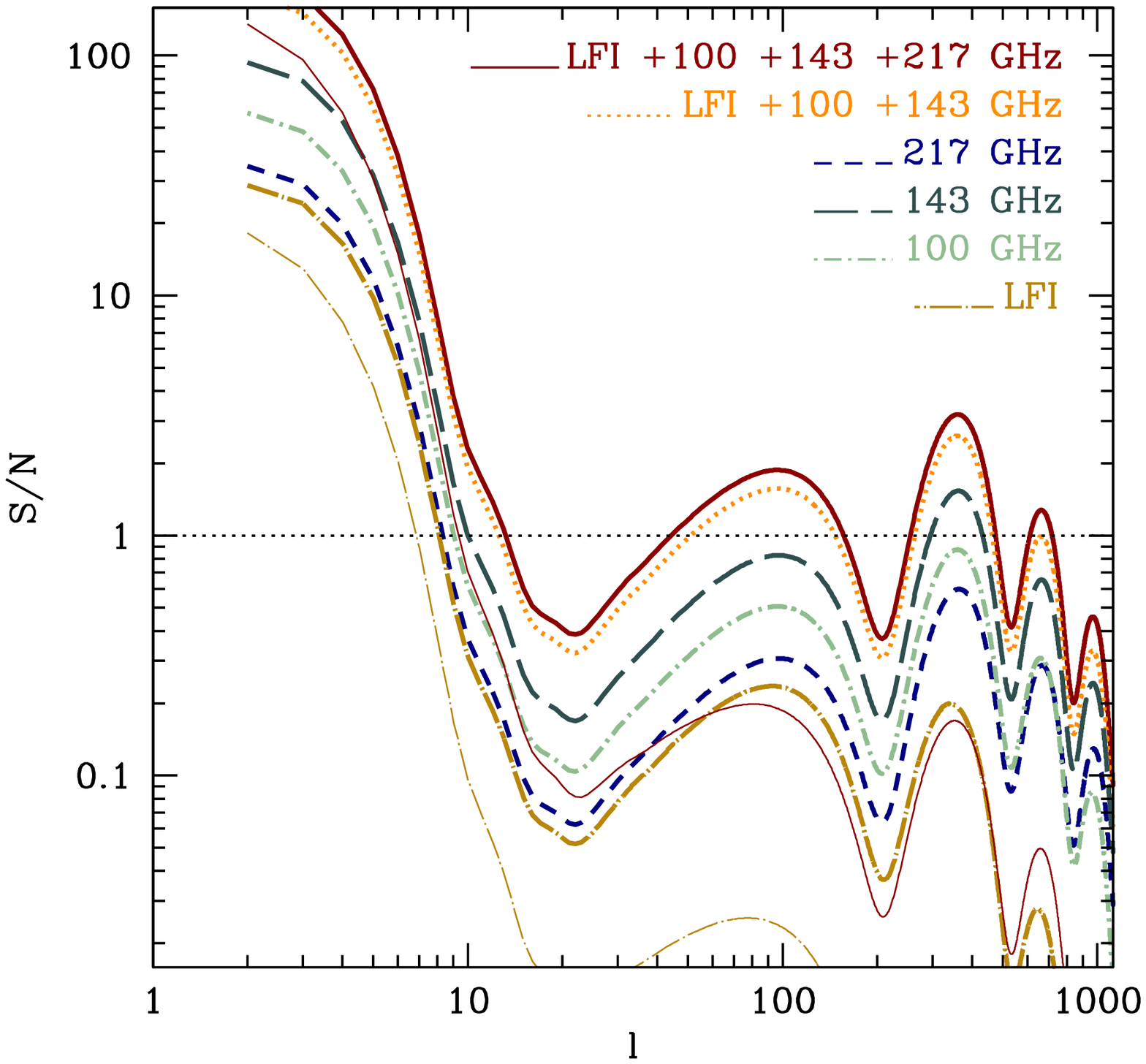}
}
\caption{Planck sensitivity to the fiducial $C_l$ spectra for
  different channels or channels combinations. Thick lines show the
  signal--to--noise ratio, thin lines show the Cosmic
  Variance--to--noise ratio. Left panel shows sensitivity to the TT
  spectrum, right panel is for the EE spectrum.}
\label{fi:s2n_T}
\label{fi:s2n_P}
\end{figure*}

Plots for power spectra in this minimal model are compared to the
noise level and cosmic variance at different frequencies in figure
\ref{fi:s2n_T}.  In temperature, all Planck channels will be ``cosmic
variance limited" (i.e. the noise equals the cosmic variance) in each
frequency channel up to angular scales $l \simeq 800$; while the
combination of all channels up to 217 GHz is cosmic variant limited up
to $ \ell \simeq 1000$.

The channels for which the signal--to--noise ratio (S/N) is unity at
the smallest scale (of $\ell \simeq 1600$) are 143 and 217 GHz, the
former having better signal to noise at larger scales and the latter
at smaller scales.  Combining all channels, the signal--to--noise ratio
becomes unity at $\ell \simeq 1800$. These figures partly depend on
the values of the reference cosmological parameters, in particular
$A_s$. WMAP5 data constrain $A_s$ at $\la 10\%$ level; the effect of
this uncertainty on the above discussion is negligible.

As for polarisation, Planck will be cosmic variance limited up to $\ell
\simeq 8-12$ (cfr with $\ell=4-5$ for WMAP5), depending on the
cosmological model considered and, in particular, on the details of
the reionisation history. The smallest scale with S/N above unity is
$\ell \simeq 800$ for a combination of channels up to 143 or 217 GHz.

In the following, we will explore how parameter constraints would vary
in the case that only a subset of frequency channels could be used for
extracting cosmological information.

\subsection{Minimal model: results on parameters}

\begin{table*}
\begin{minipage}{170mm}
\centerline{
\begin{tabular}{@{}lcccccccc}
\hline
                    &70 - 143           &30 - 143  &70 - 100 &30 - 100 &70 - 143T &30 - 143T &70 - 217 &30 - 217 \\
\hline
$\omega_{\rm b}$    &$1.6 \times 10^{-4}$  & 1.00   & 1.50     & 1.48   & 1.20     & 1.18     &  0.87      &  0.86 \\
$\omega_{\rm c}$    &$1.4 \times 10^{-3}$  & 0.99   & 1.37     & 1.33   & 1.23     & 1.19     &  0.90      &  0.90 \\
$\theta  $          &$3.1 \times 10^{-4}$  & 1.00   & 1.67     & 1.64   & 1.24     & 1.23     &  0.86      &  0.86 \\
$\tau    $          &$4.8 \times 10^{-3}$  & 0.97   & 1.25     & 1.17   & 1.23     & 1.15     &  0.95      &  0.92 \\
$n_{\rm s}     $    &$4.0 \times 10^{-3}$  & 0.99   & 1.51     & 1.47   & 1.18     & 1.14     &  0.87      &  0.87 \\
${\cal A}_{\rm s}$  &$9.5 \times 10^{-3}$  & 0.99   & 1.24     & 1.16   & 1.23     & 1.15     &  0.95      &  0.92 \\
$r$                 &$ <0.03 $             & 0.90   & 2.19     & 1.75   & 2.15     & 1.73     &  0.84      &  0.76 \\
$\sigma_8$          &$6.7 \times 10^{-3}$  & 0.99   & 1.31     & 1.25   & 1.23     & 1.17     &  0.92      &  0.91 \\
$H_0$               &$6.9 \times 10^{-1}$  & 0.99   & 1.43     & 1.39   & 1.23     & 1.19     &  0.88      &  0.88 \\
\hline
\end{tabular}
}
\caption{Error estimates for different combinations of Planck channels
  for the reference $\Lambda$CDM + tensors model. For all parameters
  we report the standard deviation for the marginalised distribution,
  except for $r$ for which we report the upper $95\%$ confidence
  limit. Second column: estimates assuming cleaning of $70$, $100$ and
  $143$ GHz channels. Values shown are actual errors. Columns 3--8
  shows estimates for different combinations of channels normalised to
  the values of column 2, when a frequency is followed by a T we only
  consider temperature data for that channel. }
\label{tab:base70}
\end{minipage}
\end{table*}

A great advantage of multi--frequency experiments is to be able to use
frequency information to subtract foreground contributions from the
maps and combine the information of different channels to derive
cosmological parameters.  In the case of WMAP5, for instance, a great
deal of leverage in parameter estimation has come from the ability to
use the Ka band in polarisation on large scales.
 
Similarly, Planck's performance will depend upon the ability to clean
foregrounds from the largest number of channels. Planck will be able
to capitalise on the existing WMAP data for its lowest frequency
channels, but being an experiment with higher sensitivity it may need
to use its own channels to perform the job fully.  The most crucial
concerns, of course, arise regarding polarisation, which is currently
poorly measured.  As there is consensus on the fact that polarised
foregrounds are minimal around 70 GHz
\citep[e.g.][]{samtleben:07,gold:08}, we will consider as a minimal
assumption that at least the 70 GHz channel will be cleaned.  WMAP5
was not able to use the 95 GHz channel in polarisation due to
potential dust contamination, however, thanks to the greater frequency
coverage of Planck, we will assume that the 100 GHz channel will also
be available to extract cosmological information.  Starting from this
minimal configuration ($70 + 100$ GHz), we will then explore the
effect of adding information from higher and lower frequencies, in
temperature only or both in temperature and polarisation.  The results
are summarised in table \ref{tab:base70}, where we quote the estimated
errors on different parameters with respect to the case for the
combination of channels $70 + 100 + 143$ GHz.  This frequency
combination is, in fact, already providing most of the information, so
that adding all LFI frequencies and the 217 GHz channels would improve
the results by only $10\%$ in all parameters.  The higher frequency
channels would add even less, due to their higher noise level.  Not
being able to use the 143GHz channel in polarisation increases errors
by about $15-25\%$ ($n_{\rm s}$ being the least impacted), except for
$r$ in which case the upper limit roughly doubles. This problem is
mitigated if all LFI channels are present.  Not being able to use the
143 GHz channel at all, however, would degrade the cosmological
parameter determination by 20--50\% on most parameter and a factor 2
on $r$. Also in this case, using all LFI gives $\sim 10\%$ improvement
on $\tau$ and $A_s$ while other parameters are affected at the percent
level.

\begin{figure*}
\centerline{
\includegraphics[width=6.cm]{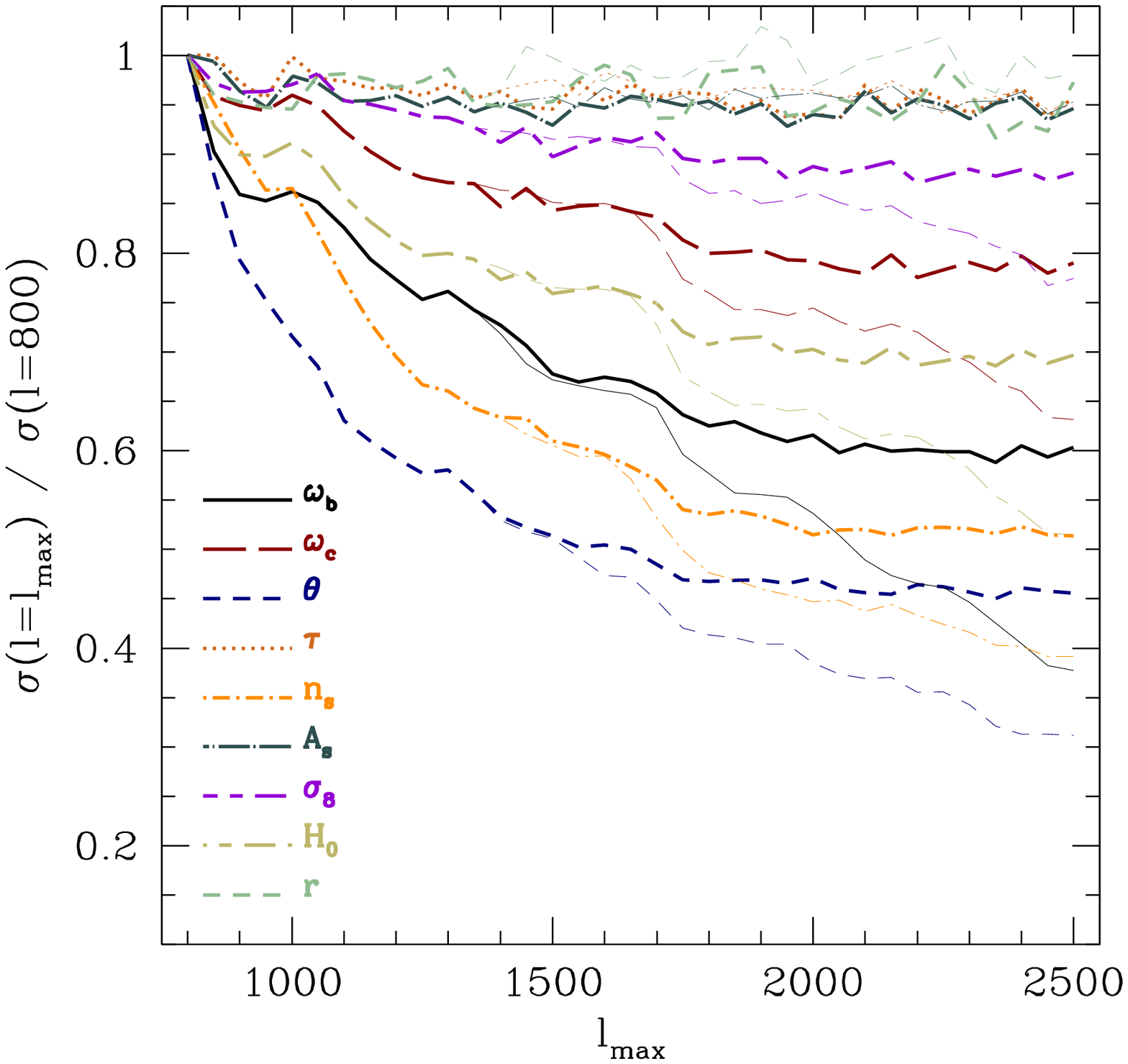}
\includegraphics[width=6.1cm]{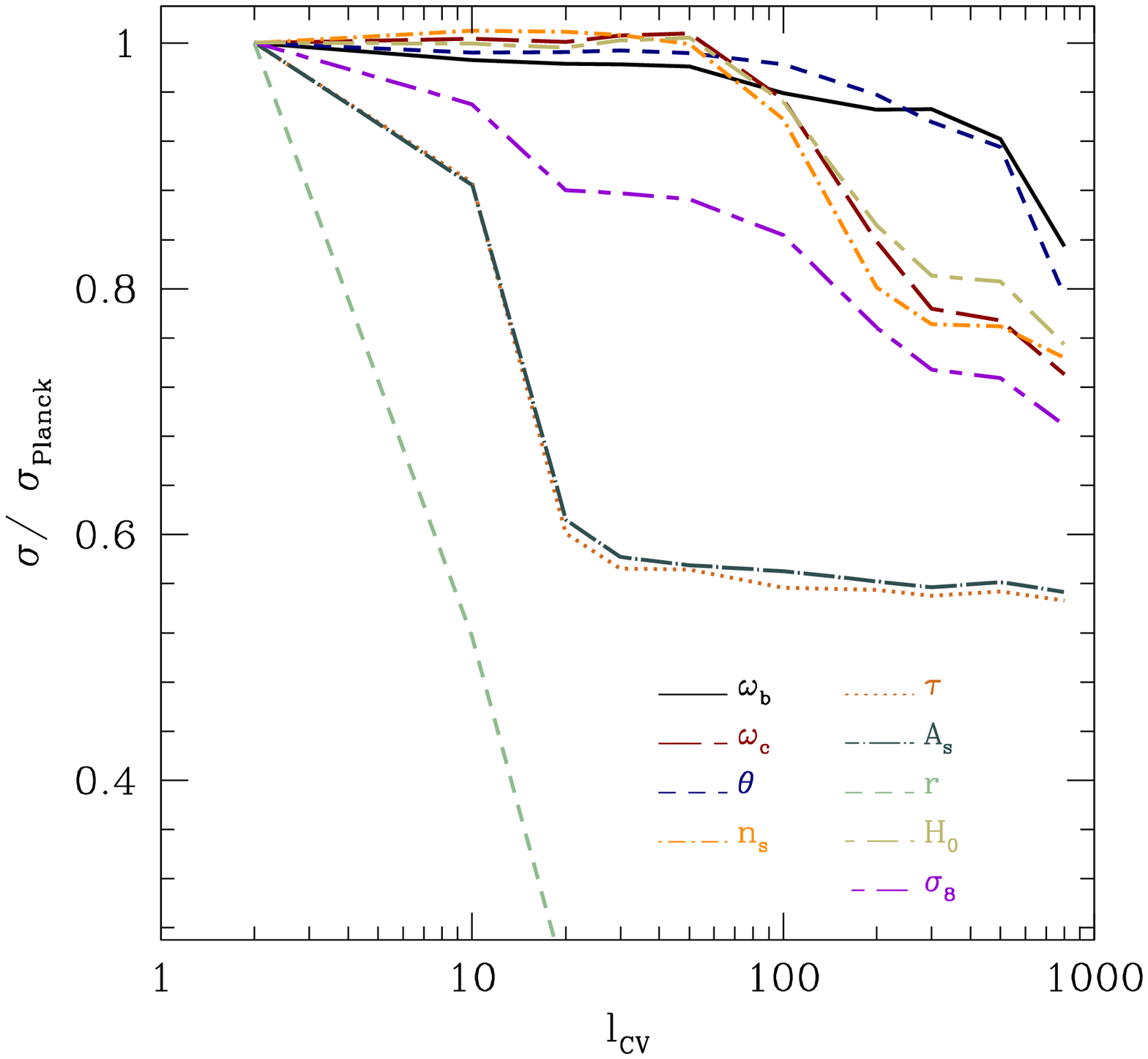}
}
\caption{{\it Left:}Sensitivity to the different parameters as a
function of the maximum multipole considered in the analysis. Plot
shows error estimates for the various parameters as a function of
$\ell_{\rm max}$ considered in the analysis, normalised to the error
for $\ell_{\rm max} = 800$.  Heavy lines are for the combination of
the $70, 100, 143$ and $217$GHz Planck channels, thin lines are for an
ideal experiment with cosmic variance temperature measurements and
polarization sensitivity equal to the Planck configuration considered
here. It is clear that up to $\ell_{\rm max} \sim 1500$ a CVL
experiment would not offer a significant advantage over Planck. {\it
Right :} Impact of higher sensitivity polarization data on parameter
constraints. Plot shows parameter estimates for an ideal experiment
with CVL polarization data up to $\ell_{\rm CV}$ and sensitivity equal
to Planck $70 - 217$ GHz channels for $\ell \ge \ell_{\rm CV}$. Error estimates
are normalised to Planck estimates, corresponding to $\ell_{\rm CV} = 2$.}
\label{fi:hybrid_lmax}
\end{figure*}

This simplified analysis summarises the information contained at
all scales. It is indeed the case that certain foregrounds are
a concern only for either temperature or polarization, or only on
some spatial scales. One such example is residual point sources,
which may spoil the use of small scale temperature fluctuations.
It is therefore interesting to ask how much of the information comes
specifically from temperature measurements and which are the relevant
scales in determining each parameter. In order to address this
question, we considered the 70 - 217 GHz channels combination and
evaluate how much each error estimate decreases as a function of the
maximum multipole $\ell_{\rm max}$ considered in the analysis. Results
are compared to the case of an ideal experiment with cosmic variance
limited (CVL) temperature measurements over the whole range of
multipoles considered, and polarization sensitivity equal to that of
the selected Planck channels. The output of this analysis is reported
in figure \ref{fi:hybrid_lmax}.

In principle, for the minimal cosmological model considered here,
Planck will recover all the relevant information on parameters encoded
in the CMB power spectra up to scales $\ell \simeq 1500$, after which
differences with a CVL experiment start to show. The parameters whose
determination is most improved when temperature information on scales
above $\ell=800$ is included are $n_{\rm s}$, $\omega_{\rm b}$ and $\theta$ (by
about 30-40\%) while $r$, $A_s$ and $\tau$ are basically unaffected.

While Planck will perform, in temperature, as well as a CVL experiment
up to $\ell \simeq 1500$, a greater room for improvement is left for
polarization measurements. This improvement may come either if Planck
keeps functioning after the 14 months required to complete 2 full sky
surveys, or by other ground--based or balloon--borne experiments, like
the currently planned SPIDER~\citep{crill:08}. SPIDER will cover $\sim
50 \%$ of the sky, with a polarization sensitivity higher than Planck,
but with a lower angular resolution. It is then interesting to
consider how the combination of Planck temperature measurements with
better polarization data will affect cosmological parameter
estimation. To asses this point we consider an ideal experiment with
temperature information equal to the combination of the $70 - 217$ GHz
Planck channels and CVL polarization measurements up to a multipole
$\ell_{\rm CV} $ ranging between $\ell = 10$ and $\ell = 800$. Above
$\ell_{\rm CV}$ polarization sensitivity is equal to that expected
from Planck. Let us also point out that CVL determination of $E$--mode
polarization above $\ell$ of a few hundreds will likely require the
next generation of space CMB mission, like the recently proposed
EPIC \citep{bock:08}. However, if such a mission does not have high angular
resolution, we do not expect a significant improvement on Planck
temperature data, which are effectively CVL for $\ell \la 1000$.

The right panel of figure~\ref{fi:hybrid_lmax} shows the improvement
on parameter constraints over Planck $70 -217$ GHz, as a function of
$\ell_{\rm CV}$. As expected, a CVL determination of even the first 10
polarization multipoles would significantly improve the constrain on
$r$. Results shown refer to a fiducial value $r = 0.05$, for which
$\ell_{\rm CV} = 10$ would improve Planck constraints by a factor
$\sim 2$; the improvement would be even more relevant for lower
tensor--to--scalar ratios. While the large angle $B$--mode spectrum is
expected to be at most comparable to Planck noise, the $E$--mode
spectrum will be measured with essentially a CVL accuracy up to $\ell =
8-9$, so that $\ell_{\rm CV} = 10$ will provide only a modest, $\sim
10 \%$, improvement in the determination of $\tau$. This improvement
will increase to $\sim 40\%$ if CVL data extend up to $\ell \sim 20$;
above this value the polarization $C_l$ are mostly independent of
$\tau$ for the class of sharp reionisation history considered here,
and further improvements on sensitivity do not lead to significant
improvements on determination of the optical depth. However, the
multipole range $20 - 50$ contains information on reionisation for
non-minimal reionisation histories, and improving determination of
this part of the spectra can significantly increase constraints on
these models. Due to the $\tau - A_s$ degeneracy, the normalisation of
the scalar power spectrum improves in a similar manner, which in turn
affects the determination of $\sigma_8$. A second set of parameters,
including $\omega_{\rm c}$, $n_{\rm s}$ and $H_0$, shows a $\sim 15 \%$
improvement for $\ell_{\rm CV} = 100 - 200$, increasing to $ \sim 25
\%$ for $\ell_{\rm CV} = 800$. A third group of parameters, $\theta$
and $\omega_{\rm b}$, shows only a modest improvement for CVL data extending
to the first 500 multipoles, and a $\sim 20 \%$ improvement for
$\ell_{\rm CV} = 800$.

In what follows, whenever not specified, we assume the frequency
combination 70 + 100 + 143 GHz and include in the analysis multipoles up
to $\ell_{\rm max} = 2500$, even if for Planck multipoles $\ell \ga 2000$
only have a (few) percent level impact on parameter determination.

\subsection{ The tensor--to--scalar ratio determination}
\label{sec:r}

As it appears from table \ref{tab:wmap}, one parameter whose
determination Planck is expected to greatly improve is the
tensor--to--scalar ratio $r$. For a fiducial model with negligible
tensor contribution, $r \la 0.01$ corresponding to {\it small--field}
inflation models \citep[for a review of inflationary physics,
  see][]{lyth:99}, using the $70 - 143$ GHz Planck channels we expect
an upper limit $r \la 0.03 (95\% {\rm c.l.})$, tightening WMAP5
constraints by about 1 order of magnitude.  Without the polarization
information of the 143 GHz channel, this constraint degrades by more
than a factor 2, while adding the 217 GHz channel would improve this
figure by $\sim 15 \%$. If it proves possible to use all the LFI
channels for cosmological parameter estimation, the upper limit would
improve by $ \sim 25 \%$ if the 143 GHz polarization is not included
in the analysis, or by $\sim 10 \%$ otherwise.

\begin{figure*}
\centerline{
\includegraphics[width=6.cm]{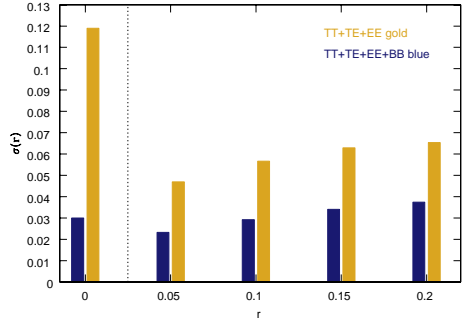}
}
\caption{Impact of $B$--modes detection on the determination of the
  tensor--to--scalar ratio, $r$, for different fiducial values of
  $r$. For $ r > 0$ bars show the expected $1\sigma$ error on the
  tensor--to--scalar ratio including (blue) or excluding (gold)
  $B$--modes in the analysis. For a fiducial $r= 0$ bars show the
  upper $95 \%$ c.l. instead. If $B$--modes information is not
  available, error on $r$ increase by a factor $\sim 2$. Notice that
  for $r \la 0.10$, the corresponding marginalised distribution is
  markedly non--Gaussian and quoting the standard deviation does not
  properly characterise the $1\sigma$ confidence interval. Results
  shown are for the $70 - 143$ GHz channels combination.}
\label{fi:Bmodes_r0}
\end{figure*}

For a non--negligible, i.e. $r \ga 0.05$, contribution by
gravitational waves, Planck will be able to put a lower limit on $r$
with a confidence level $\ga 95 \%$ (see figure \ref{fi:Bmodes_r0}).
It is then interesting to explore to what extent results depend on
the frequencies considered. Results for a fiducial cosmological model
with $r = 0.05$ are given in table \ref{tab:tens}.  A combination of
the 70--143 GHz channels allows $r$ to be constrained with an accuracy of
$\sigma_r \sim 0.023$. The uncertainty on $r$ increases by $\sim 50\%$
without the polarization information from the 143 GHz channel,
however, since the marginalised distribution is markedly non--Gaussian
even in this case it is possible to claim a $2 \sigma$ detection of $
r > 0$. Adding the $30$ and $40$ GHz channels would improve
constraints on $r$ by $10-15 \%$. Notice that these results were
obtained by fixing the tensor spectral index to $n_{\rm T} = 0$ (see below
for further discussion).

The BB spectrum is a unique signature of tensor perturbation, however
it is also a weaker signal than the EE and TE spectrum; it has not yet
been observed, and is potentially more affected by foreground
residuals.  Simulations of polarised foregrounds removal strategies
\citep{betoule:08} suggest that for Planck residual foreground
contamination would increase the uncertainty on $r$ by $\sim 30 \%$,
for a fiducial value $r = 0.10$. Thus, we expect foreground cleaning
to have an impact on the uncertainty on $r$ which is (at least)
comparable to that of the instrumental noise. However, a detailed
knowledge of polarised foregrounds is still lacking, and a correct
assessment of Planck capabilities of measuring the $B$--mode $C_\ell$
will require the actual data.

For these reasons, it is sensible to ask how much of the constraining
power resides in the BB measurement and to what extent the constraint
on $r$ would be weakened if instead we were only able to measure TT,
EE and TE. The results are summarised in figure \ref{fi:Bmodes_r0}
for $r=0, 0.05, 0.1, 0.15, 0.2$, and a fiducial set of parameters as
above. In the case of a high value of $r$ that Planck might possibly
detect, not being able to use the BB spectrum would imply a doubling
of the error bars, and a $3 \sigma$ detection would be possible only
for values $r \ga 0.15 -0.20$, in agreement with \cite{knox:94}.

It is then clear that a significant portion of Planck's capabilities
of detecting tensor modes comes from the low--$\ell$ part of the BB
spectrum. When dealing with real data, contamination by the Galaxy, as
well as component separation and point sources subtraction, will
require masking of parts of the sky. In this case, the correct shape
of the likelihood function is no longer given by equation
~\ref{eq:wishart}, and for arbitrary cuts it is not possible to write
an analytical expression for the likelihood in harmonic space. As
discussed above, we assumed here that it is possible to account for
sky cuts simply by multiplying the Wishart distribution by a factor
$f_{\rm sky}^2$. To test that this approximation does not
significantly affect our determination of $r$, we compare it with the
exact likelihood function evaluated in pixel space \citep[we refer the
  interested reader to, e.g.,][for the relevant
  expressions]{tegmark:01}. Evaluating the full Planck likelihood in
pixel space is not numerically feasible, since it would require the
inversion of a matrix with a side of $O(10^7)$ elements. However, we
focus here on the determination of $r$ which depends mainly on the
$\ell \la 30$ multipoles, so that we can work with low resolution
maps. Using the HEALPix package\footnote{http://healpix.jpl.nasa.gov/}
\citep{gorski:05}, we generate temperature and polarization maps at a
resolution $N_{\rm side} =16$, corresponding to a pixel size of $\sim
3.5°$, including contribution by multipoles up to $\ell = 32$. We add
a white noise corresponding to that expected for the $70 - 140$ GHz
channels; due to the low resolution of the maps, the beam finite
resolution is not relevant. From each map we remove the pixels with
centers within $10°$ of the equator. Using the remaining pixels,
corresponding $f_{\rm sky} = 0.8125$, we evaluate the exact likelihood
for $r$, with other parameters fixed to the fiducial value. We compare
this likelihood to the one evaluated with equation \ref{eq:wishart},
using as fiducial spectrum the full--sky spectrum of the CMB + noise
map. Both in pixel and harmonic space, we include in the likelihood
evaluation TT, TE, EE and BB multipoles up to $\ell = 32$.  We repeat
this test for two fiducial values $r = 0$ and $r=0.10$, for each value
we perform 50 sky realization. For $r = 0$ we compare the 95\% upper
limits of the pixel and harmonic based likelihood, and find that in
average the approximate likelihood we use throughout this work
overestimates the correct result by $\sim 4 \%$. For $r= 0.10$, we
instead compare the variance of the two sets of distributions; using
the approximate likelihood overestimates the correct results by $\sim
11 \%$.  As expected, when considering single realizations, the
estimate of $r$ obtained with the rescaled Wishart distribution are
different from that obtained in pixel space, as the former likelihood
peaks at the full--sky power spectrum, while the latter correctly
takes into account only the observed portion of the sky. On average
this discrepance is $\sim 0.35 \sigma_r$. Elsewhere in this work we
build the mock data using the theoretical power spectrum, instead of
considering realizations, therefore there is no bias in the
corresponding results. Since $r$ is not significantly degenerate with
the remaining parameters, with the exception of $n_{\rm T}$, we expect
that these conclusions hold also when fitting for the remaining
cosmological parameters, and that results of this paper do not
significantly depend on the sky coverage of actual data.

\begin{table}
\centerline{
\begin{tabular}{lcccc}
\hline
                    & 70 - 143 & 70 - 100 & 70 - 143T & 70 - 217  \\
\hline
$\omega_{\rm b}$          & $1.6 \times 10^{-4}$  &  1.55   & 1.20       &  0.89      \\
$\omega_{\rm c}$          & $1.4 \times 10^{-3}$  &  1.38   & 1.21       &  0.88      \\
$\theta  $          & $3.1 \times 10^{-4}$  &  1.69   & 1.25       &  0.87      \\
$\tau    $          & $4.9 \times 10^{-3}$  &  1.24   & 1.24       &  0.94      \\
$n_{\rm s}     $          & $4.1 \times 10^{-3}$  &  1.58   & 1.17       &  0.85      \\
${\cal A}_{\rm s}$        & $9.8 \times 10^{-3}$  &  1.22   & 1.23       &  0.95     \\
$r$                 & $2.3 \times 10^{-2}$  &  1.52   & 1.50       &  0.85      \\
$\sigma_8$          & $6.9 \times 10^{-3}$  &  1.29   & 1.22       &  0.91     \\
$H_0$               & $7.0 \times 10^{-1}$  &  1.45   & 1.21       &  0.87      \\
\hline
\end{tabular}
}
\caption{As table~\ref{tab:base70}, but assuming a non--zero tensor
  amplitude, $r =0.05$.  In this case for $r$ we quote the standard
  deviation as done with all the other parameters.}
\label{tab:tens}
\end{table}

\subsection{Beam degradation}

\begin{figure*}
\centerline{
\includegraphics[width=6cm]{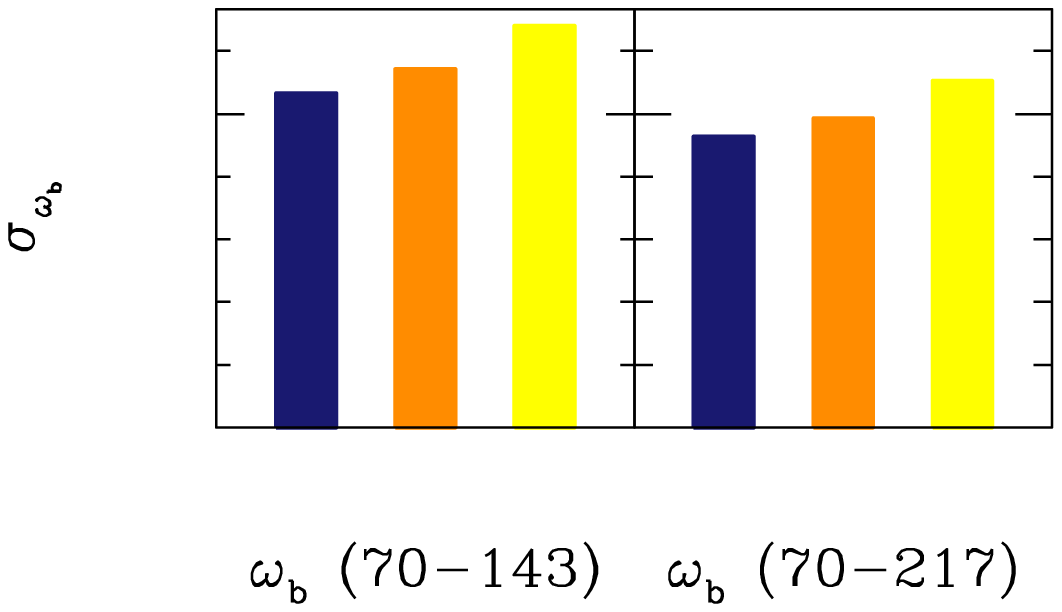}
\includegraphics[width=6cm]{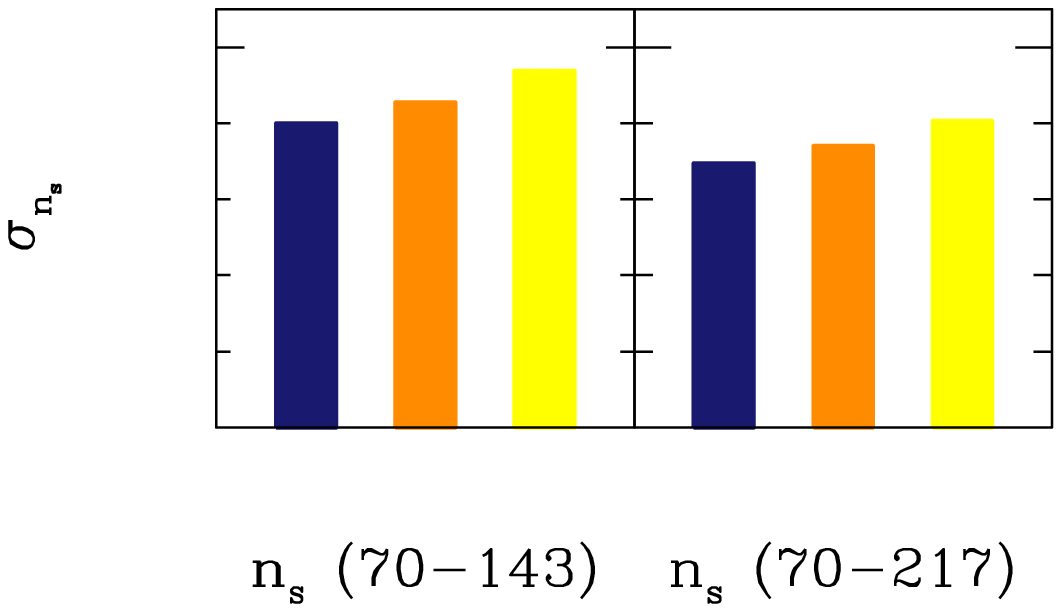}
\includegraphics[width=6cm]{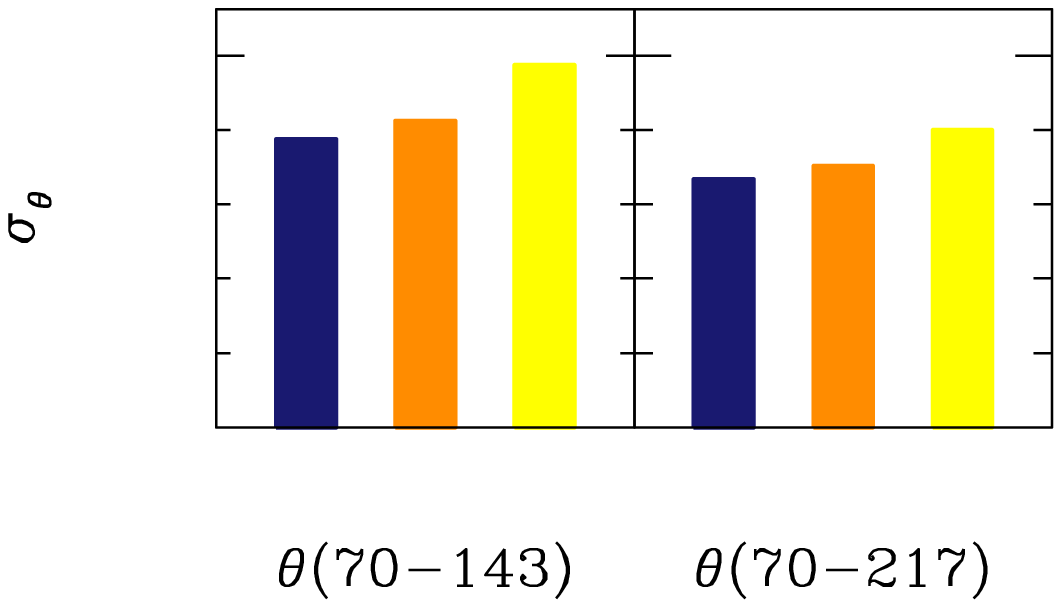}
}
\caption{Impact of small angular scales on parameter determination.
  Histograms shows the uncertainties on $\omega_{\rm b}$, $n_{\rm s}$
  and $\theta$ for nominal beams (left bar), beam enlarged by $10\%$
  (middle) or by $30 \%$ (right). For each parameter, the left
  panel refers to the $70 - 143 $GHz channels, the right panel is
  for $70 - 217 $ GHz channels.}
\label{fi:beams}
\end{figure*}

Planck results will come from a combination of high sensitivity and
fine angular resolution. While a comparison of Planck with a CVL
experiment gives an assessment of the relevance of high sensitivity, a
crude test of the impact of angular resolution can be made by
comparing Planck's expected performance with that of an ideal experiment
with {\it ad--hoc} increased beam FWHM at all frequencies by either
10\% or 30\%. 

Table \ref{tab:beams} reports the parameter uncertainties for such an
experiment, relative to the Planck expected performances quoted in
table \ref{tab:base70}.  An increase of 10\% in FWHM worsen the
constraints on most parameters by $ \sim 5\% - 10\%$, while degrading
the beams by 30\% increases error estimates by $ \sim 10\% - 20\%$,
for our basic configuration. As shown in figure \ref{fi:beams}, some
parameters, most noticeably $n_{\rm s}$, $\omega_{\rm b}$ and
$\theta$, gain constraining power from the measurements of high--$\ell$ TT
power spectrum.

\begin{figure*}
\centerline{
\includegraphics[width=10.cm]{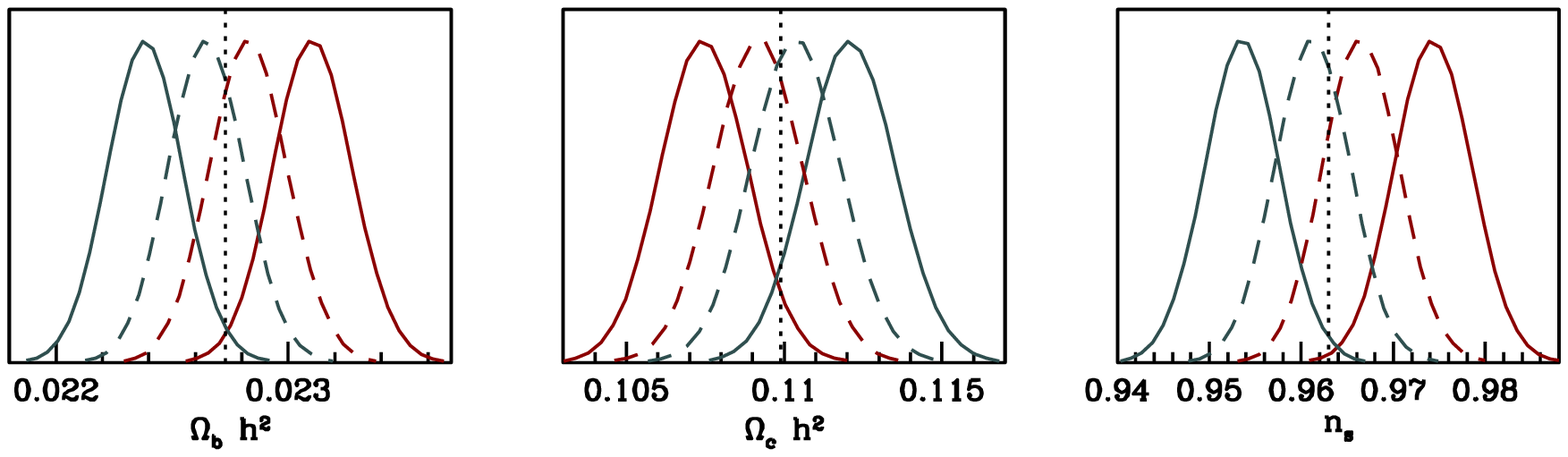}
}
\caption{Effect of an incorrect beam characterisation on parameter
estimation for Gaussian beams. Figure shows the bias in the recovered
parameter resulting by underestimating (red lines) or overestimating
(green lines) the actual beam FWHM by $0.05 \%$ (dashed lines) or $0.20
\%$ (solid).  }
\label{fi:beams2}
\end{figure*}

Real beams can be affected by a number of systematic effects,
e.g. deviations from Gaussianity or asymmetries. A full assessment of
these effects will require in--flight beam calibration, and is outside
of the scope of this paper. Here instead, we perform a simple test of
beam impact on parameter estimation by assuming that the Gaussian FWHM
used in the analysis is different (either bigger or smaller) by the
actual FWHM by $0.05 \%$ or $0.20 \%$, for all frequencies
considered. The major impact of a beam mismatch is a bias on most
cosmological parameters, as shown in figure~\ref{fi:beams2}. Even an
error of $0.05 \%$ in the assumed beams leads to a bias of $\sim ~0.5
\sigma$ on most parameters, increasing to more the $\sim 2 \sigma$ for
a misestimate of the FWHM by $0.20 \%$. For Gaussian beams, the beam
transfer function depends on a single $\ell$. Thus, if the beam
mismatch becomes relevant only at scales $\ell \ga 1800-2000$, it is
expected to affect parameter estimation at the few percent level (see
figure~\ref{fi:hybrid_lmax}). Real systematics, however, will also
introduce correlations between different multipoles so in general it
will not be possible to reduce these effects by simply excluding this
range of $\ell$'s.

\begin{table}
\centerline{
\begin{tabular}{lcc}
\hline
                    & FWHM = 1.1 FWHM$_{\rm Planck}$ & FWHM = 1.3 FWHM$_{\rm Planck}$  \\
\hline
$\omega_{\rm b}$          & 1.07   &  1.20 \\
$\omega_{\rm c}$          & 1.03   &  1.10 \\
$\theta  $          & 1.08   &  1.25 \\
$\tau    $          & 1.00   &  1.01 \\
$n_{\rm s}     $          & 1.05   &  1.16 \\
${\cal A}_{\rm s}$        & 1.00   &  1.00 \\
$r$                 & 0.99   &  1.00 \\
$\sigma_8$          & 1.03   &  1.06 \\
$H_0$               & 1.05   &  1.14 \\
\hline
\end{tabular}
}
\caption{Error estimates for an experiment with sensitivity equal to Planck
70-143GHz channels and larger beams. Values are quoted relative to Planck
estimates reported in table~\ref{tab:base70}.}
\label{tab:beams}
\end{table}

\subsection{Away from the minimal model}

After having explored Planck's performance for the minimal cosmological
model, we now broaden the parameter space and investigate performance
for a series of less minimal but still quite general cosmological
models.  These will include a running spectral index,
neutrino--related parameters, curvature and quintessence. For these
models, we will explore frequency dependence and degeneracies.

\subsubsection{Dark Energy and Cosmic Curvature}
\label{sec:geo}

In most cosmologies, the size of the sound horizon at last scattering
provides a standard ruler at a redshift of $z_{\rm LS} \simeq
1090$. Indeed, most information on the geometry of the Universe
present in the CMB power spectra, can be summarised in terms of the
redshift of photon decoupling, $z_*$, and of two distance ratios, the
acoustic scale $\ell_A$ and the shift parameter $R$
~\citep{elgaroy:07,mukherjee:08,komatsu:08}. In principle this allows
for a substantial compression of information without a significant
loss of accuracy, especially when combining CMB data with external
measurements. This approach however cannot be adopted when
simultaneously fitting for the entire set of cosmological parameters
we are considering.

The geometrical degeneracy limits how the accuracy on the spectra
translates in accuracy on the cosmological parameters defining the
geometry of the Universe or dark energy properties and current
constraints on these parameters rely heavily on external
information. We wish to investigate here to what extent this will be
true for Planck as well.  In particular, we will consider cosmic
curvature, $\Omega_{\rm K}$, and a dark energy component with a
constant equation of state, $w$.

Constraining the geometry of the Universe to high precision provides a
weak, but important probe of inflation.  Generic inflation models tend
to produce a Universe completely flat except for super-horizon
curvature fluctuations at the level of $\Omega_{\rm K}\sim 10^{-5}$.
Although this high level of precision is likely beyond currently
envisaged experiments a positive detection of curvature at a level
higher than this would be a very interesting result.  This would not
necessarily invalidate the inflationary paradigm, since open inflation
models \citep{linde1995} can produce $\Omega_{\rm K}\la 10^{-3}$.
Measurements at this level would move towards constraining such
models.

\begin{figure*}
\centerline{
\includegraphics[width=6.cm]{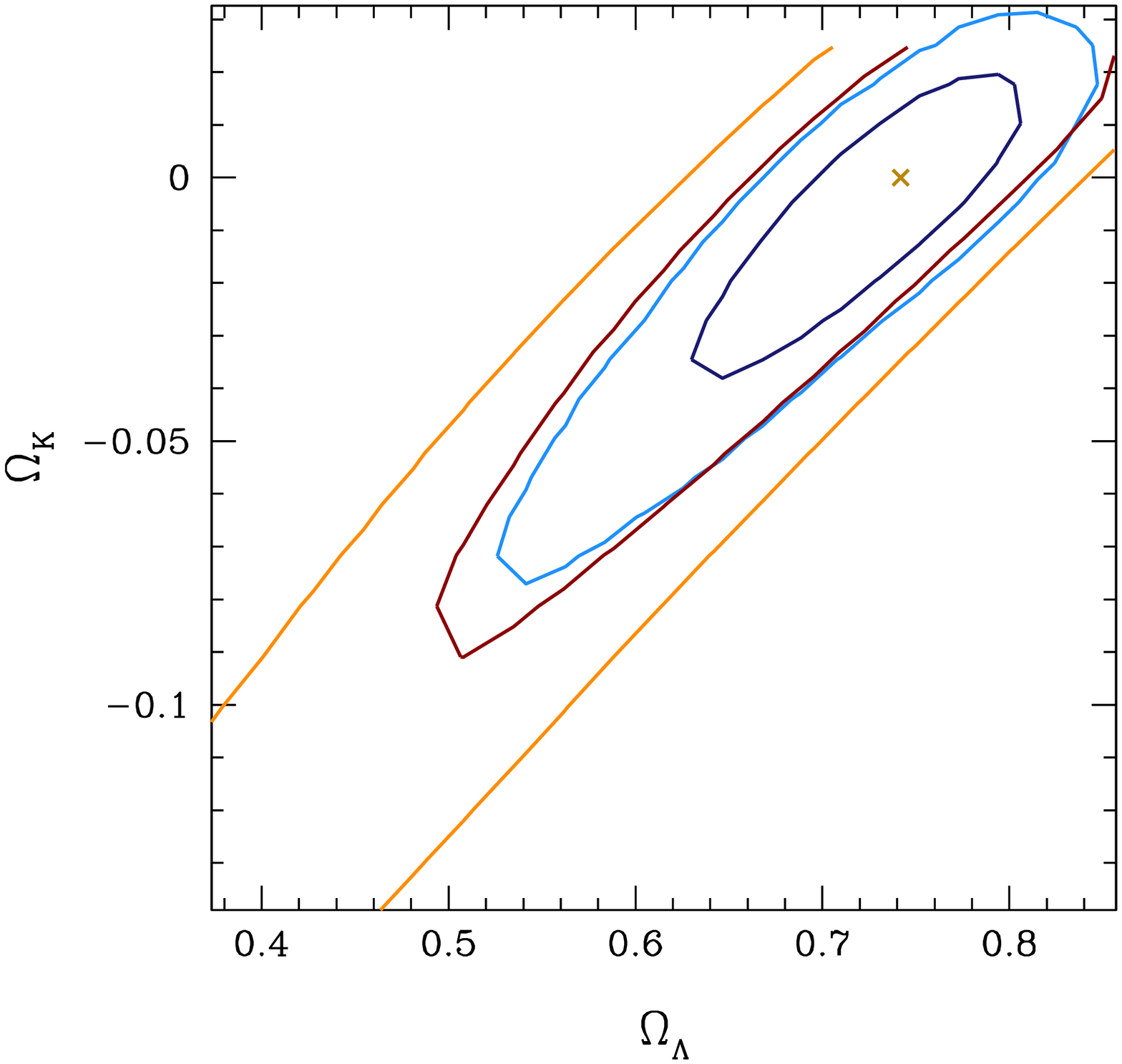}
\includegraphics[width=6.cm]{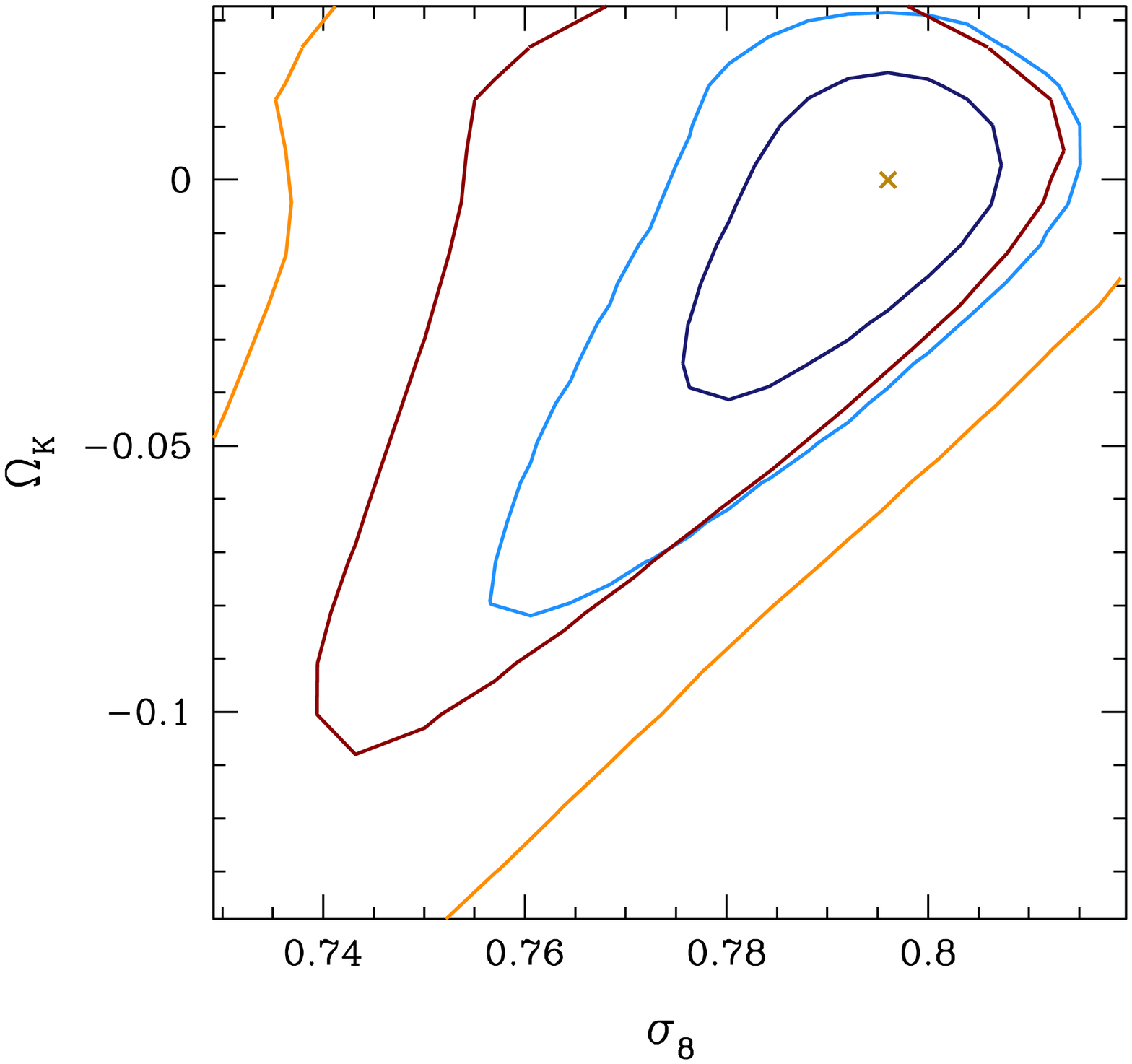}
}
\caption{Joint $1-$ and $2-\sigma$ confidence regions in the
  $\Omega_\Lambda-\Omega_{\rm K}$ (left) and in the
  $\Omega_\Lambda-\Omega_{\rm K}$ (right) planes. Plots compare
  results for the $70 - 143$GHz channels (blue/cyan lines) with those
  for simulated WMAP5 data (red/orange). Crosses mark the input value
  of the parameters. Planck will significantly reduce the degeneracies
  present in WMAP5 data.}
\label{fi:omk_oml}
\end{figure*}

Table~\ref{tab:open70} reports results when the parameter space is
expanded to include $\Omega_{\rm K}$. Our default Planck configuration
will constrain curvature with an accuracy of $\sigma_{\Omega_{\rm K}}
\sim 0.023$, compared to an accuracy of $\sigma_{\Omega_{\rm K}} \sim
0.050$ for the WMAP5--like case (see
figure~\ref{fi:omk_oml}). Allowing for curvature degrees of freedom in
the analysis increases the error on $H_0$ by more than an order of
magnitude and almost doubles the error on $\sigma_8$. As expected,
other parameters are not significantly affected.

\begin{table}
\centerline{
\begin{tabular}{lcccc}
\hline
                    & 70 - 143 & 70 - 100 & 70 - 217  \\
\hline
$\omega_{\rm b}$          &$1.7 \times 10^{-4}$  &  1.48   &   0.85        \\
$\omega_{\rm c}$          &$1.4 \times 10^{-3}$  &  1.39   &   0.91        \\
$\theta  $          &$3.1 \times 10^{-4}$  &  1.64   &   0.85        \\
$\tau    $          &$4.9 \times 10^{-3}$  &  1.23   &   0.92        \\
$\Omega_{\rm K}$     &$2.3 \times 10^{-2}$  &  1.33   &   0.90        \\
$n_{\rm s}     $          &$4.1 \times 10^{-3}$  &  1.54   &   0.88        \\
${\cal A}_{\rm s}$        & $9.8 \times 10^{-3}$ & 1.20  &   0.93       \\
$r$                 & $0.031$          &  2.43   &   0.83        \\
$\sigma_8$          & $1.1 \times 10^{-2}$  &  1.35   &   0.91       \\
$H_0$               & $9.2     $  &  1.15   &   0.93        \\
\hline
\end{tabular}
}
\caption{As table~\ref{tab:base70}, but for a $\Lambda$CDM + tensor +
  curvature model.}
\label{tab:open70}
\end{table}

We now turn to discussing estimation of dark energy properties. The
simplest way to account for dark energy is to {\it ad--hoc} include a
cosmological constant term into Einstein's equations. A more physically
motivated alternative is provided by a self--interacting scalar field
\citep{ratra:88,wetterich:88,caldwell:98,brax:99}. A common prediction
of these models is that dark energy can be described as a perfect
fluid with a time varying equation of state, $w(a)$. In general, the
effects of dark energy on CMB spectra are well described by modelling
dark energy as a perfect fluid with constant equation of state
\begin{equation}
w = \frac{ \int da \, \Omega_w(a) w(a)}{\int da \, \Omega_w(a)}~.
\end{equation}
This approximation is not valid for models with direct dark energy --
dark matter interaction
\citep[e.g.,][]{amendola:00,huey:06,mainini:05} and may lead to
significant biases in the determination of cosmological parameters
\citep{vergani:08}. We will not consider the latter class of models
here, and parametrise dark energy with a constant equation of state,
$w$. We also assume a flat Universe.

\begin{figure*}
\centerline{
\includegraphics[width=6.cm]{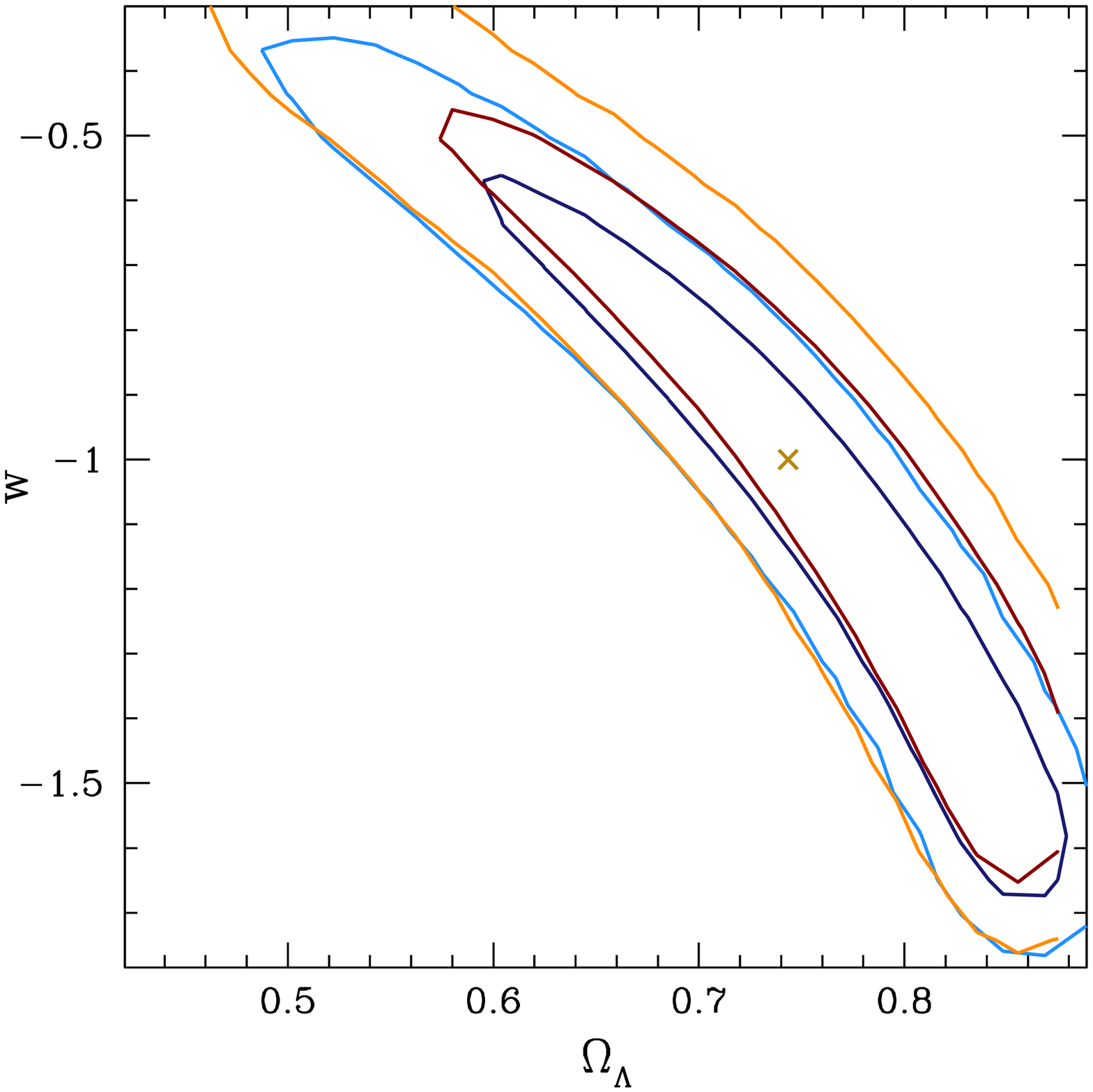}
\includegraphics[width=6.cm]{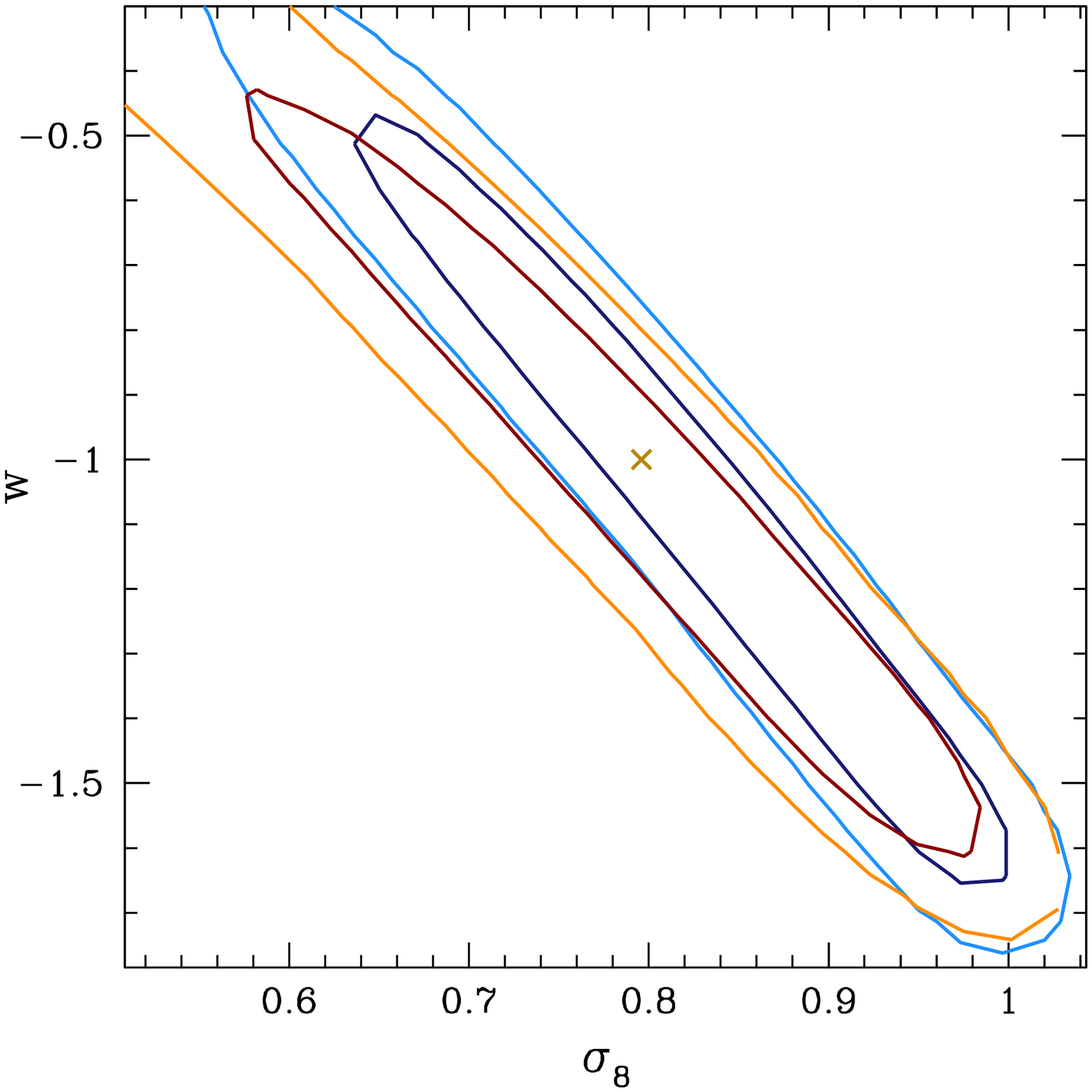}
}
\caption{ Joint $1-$ and $2-\sigma$ confidence regions in the
  $w-\Omega_\Lambda$ (left) and in the $w - \sigma_8$ (right) planes,
  for a combination of $70 - 143$ GHz Planck channels (blue/cyan
  lines) or for WMAP5 (red/orange lines). Cross marks the input value
  of the parameters. Results are dominated by intrinsic CMB
  degeneracies and by the choice of priors (especially on $H_0$).}
\label{fi:w_degs}
\end{figure*}

Error estimates for WCDM models are quoted in table~\ref{tab:w70},
while figure~\ref{fi:w_degs} compares degeneracy regions for Planck
and WMAP in the $w-\Omega_\Lambda$ and $w-\sigma_8$ planes. As with
curvature, allowing for $w \ne -1 $ significantly increases the errors
on $H_0$ and $\sigma_8$. In this case, however, estimates are not
dependent on the combination of channels used, and in addition Planck
provides only a marginal improvement over WMAP5. Thus, at Planck
sensitivities constraints are driven by the CMB internal degeneracies
and the prior used in the analysis (e.g. on $H_0$), rather than by the
data.

In summary, Planck will significantly improve over WMAP 5--year
constraints on $\Omega_{\rm K}$, but not on $w$. However, Planck will
significantly improve over WMAP5 if we are considering models with
both non--zero curvature and $w \ne -1$, or models with more than one
dark energy parameter \citep{xia:07}. In both cases, external data set
will still be fundamental \citep[also see][]{mukherjee:08}.

\begin{table}
\centerline{
\begin{tabular}{lcccc}
\hline
                    & 70 - 143 & 70 - 100 & 70 - 217  \\
\hline
$\omega_{\rm b}$          &$1.6 \times 10^{-4}$  &  1.53   &   .897        \\
$\omega_{\rm c}$          &$1.4 \times 10^{-3}$  &  1.40   &   .924        \\
$\theta  $          &$3.1 \times 10^{-4}$  &  1.65   &   0.85        \\
$\tau    $          &$4.8 \times 10^{-3}$  &  1.23   &   .945        \\
$w$                 &$3.5 \times 10^{-1}$  &  1.01   &   0.99        \\
$n_{\rm s}     $          &$4.0 \times 10^{-3}$  &  1.55   &   .886        \\
${\cal A}_{\rm s}$        & $9.7 \times 10^{-3}$ & 1.23  &   .953       \\
$r$                 & $0.030$          &  2.18   &   .819        \\
$\sigma_8$          & $1.1 \times 10^{-2}$  &  1.02   &  0.99       \\
$H_0$               & $1.3 \times 10    $  &  1.01   &   0.99        \\
\hline
\end{tabular}
}
\caption{As table~\ref{tab:base70}, $\Lambda$CDM + tensor + w
  model. Errors on $w$, $\sigma_8$ and $H_0$ do not depend on the
  channels used.}
\label{tab:w70}
\end{table}

\subsubsection{Neutrinos and relativistic degrees of freedom}

The detection of oscillations of solar and atmospheric
neutrinos has confirmed that $\nu$'s are massive particles. However,
for currently allowed values of neutrino's mass, $\sum m_\nu \la 1.3
\eV$ (95\% c.l.) \citep{komatsu:08}, neutrinos will still be relativistic at the
epoch of last scattering, so that the effective matter--to--radiation
ratio at $z_{\rm LS}$ is lower than the matter--to--radiation ratio
today. In turn, this results in a shift of the position of the first
acoustic peak due to the faster decay of the gravitational potential
around recombination, and alter the shape of the acoustic
peaks~\citep{pierpa:03,ichikawa:05,lesgourgues:06}. In addition, neutrino mass
slightly affect the expansion rate at late times.

The three standard neutrinos may not be the only free streaming
relativistic particle species present in the early Universe.
Additional relativistic degrees of freedom, usually parametrised in
terms of effective number of neutrino species, $N_{\rm eff}$, would
push the redshift of matter--radiation equality to later epochs, which
in turn affects the ratio of the heights of the first and third
acoustic peaks of CMB spectra. Moreover, extra relativistic degrees of
freedom change the expansion rate at early times, in particular
affecting the time between (standard) neutrinos decoupling and the
opening of the Deuterium bottleneck, which fixes the abundance of
light elements produced in the Big Bang Nucleosynthesis. Knowledge of
the Helium abundance, $Y_{\rm He}$, would then tighten constrain on
$N_{\rm eff}$.  WMAP5 data alone are not able to place an upper limit
on $N_{\rm eff}$, even assuming $Y_{\rm He}$ is known. With its
improved characterisation of the acoustic peaks, Planck is expected to
significantly constrain both these parameters \citep[also
  see][]{ichikawa:08}.

Here, we discuss the determination of neutrino masses separately from
that of the effective number of relativistic species and Helium
abundance. We show how these constraints depend on the frequencies
used, and how errors on the non--neutrinos related parameters increase
with respect to results for a $\Lambda$CDM model. The fiducial
cosmological model does not include massive neutrinos, $f_\nu = 0$, or
additional massless species, $N_{\rm eff} = 3.04$, and $Y_{\rm He} =
0.24$.

When expanding the parameter space to include $\nu$'s mass, we assume
that the three mass eigenstates are completely degenerate and
parametrise the neutrino contribution to the total dark matter energy
density by $ f_\nu \equiv \Omega_{\nu,0} / \Omega_{\rm dm, 0} $ with
$\Omega_{\rm dm, 0} \equiv \Omega_{c,0} +\Omega_{\nu,0}$. Results for
the $70-143$ GHz are reported in table~\ref{tab:numass}. We find an
upper limit $f_\nu < 0.072$ ($95\%$ c.l.), which translates into a
constraint on the sum of $ \nu$ 's masses $\sum m_\nu < 0.77 \, \eV$,
according to $ \sum m_\nu = 94 \, {\eV} \, \Omega_{\nu, 0} h^2 = 94 \,
{\eV} f_\nu \Omega_{\rm dm, 0} h^2$. Using only the $70$ and $100$ GHz
channels, this limit increases by $\sim 20 \%$, while if the $217$GHz
channel is included the limit improves by $\sim 5\%$\footnote{Due to
the stochastic nature of MCMC methods, the $2\sigma$ upper limit on
$f_\nu$ is determined with a 5\% accuracy.}. Allowing for a
contribution by massive neutrinos more than doubles the uncertainty on
$\omega_{\rm dm}$ and increases errors on $H_0$ by $\sim 3.5$
times. Since $\sigma_8$ is an integrated quantity with a significant
dependence on $f_\nu, \omega_{\rm dm}$ and $H_0$, these effects
combines so that the corresponding uncertainty increase by a factor
$\sim 7-8$. Errorbars on other parameters are only moderately
affected, increasing by $\sim 5 -10 \%$, implying no significant
degeneracies with $f_\nu$. When fitting for neutrino properties, we
assume a constant equation of state dark energy with $w = -1$. This
assumption does not affect our results since the $m_\nu$ - $w$
degeneracy is relevant only when combining CMB data with external
measurements such as galaxy surveys or Ly$\alpha$ forest data
\citep{hannestad:05,komatsu:08}, unless we also allow for a direct
dark energy -- dark matter interaction \citep{lavacca:08b}.

\begin{figure*}
\centerline{
\includegraphics[width=12.cm]{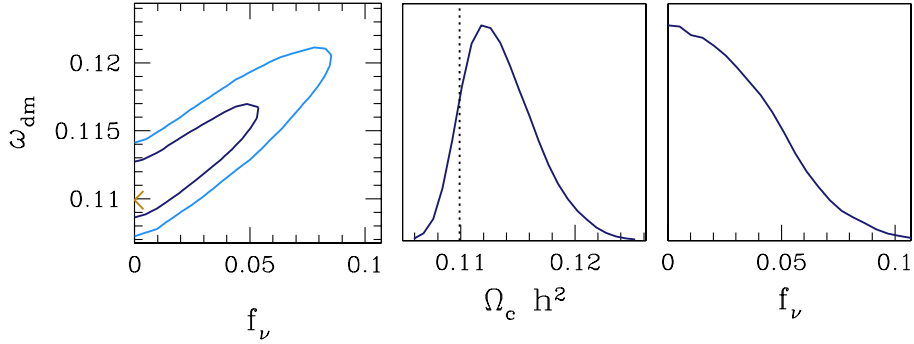}
}
\caption{{\it Left}: Degeneracy between the neutrino fraction, $f_\nu$ and
  the total dark matter density, $\omega_{\rm dm}$. {\it Middle and Right}:
  Marginalised distribution over $\omega_{\rm dm}$ and $f_\nu$,
  respectively. When marginalising over $f_\nu$, the non --
  Gaussianity in the distribution of the neutrino fractions transfers
  to $\omega_{\rm dm}$, which leads to overestimating the dark matter
  density.}
\label{fi:fnu_omc}
\end{figure*}

In addition to the increase in uncertainties, the degeneracy between
$f_{\nu}$ and $\omega_{\rm dm}$ leads to overestimate the dark matter
density by $\sim 1 \sigma$, as shown in figure~\ref{fi:fnu_omc}
\citep[also see][]{perotto:06}.

\begin{table}
\centerline{
\begin{tabular}{lcccccc}
\hline
                    & 70 - 143                 & 70 - 100 & 70 - 217 &\\ 
\hline
$\omega_{\rm b}$          &$1.7 \times 10^{-4} \, (1.06)$   & 1.54 & 0.87 \\
$\omega_{\rm dm}$    &$3.1 \times 10^{-3} \, (2.21)$   & 1.29 & 0.93 \\
$\theta  $          &$3.4 \times 10^{-3} \, (1.13)$   & 1.59 & 0.87 \\
$\tau    $          &$4.9 \times 10^{-3} \, (1.04)$   & 1.24 & 0.94 \\
$n_{\rm s}     $          &$4.3 \times 10^{-3} \, (1.08)$   & 1.54 & 0.87\\
${\cal A}_{\rm s}$        &$1.0 \times 10^{-2} \, (1.06)$   & 1.23 & 0.94\\
$f_{\nu}         $   &$ < 7.2 \times 10^{-2} $        & 1.21 & 0.95\\
$r$                 &$< 0.034 \, (1.13)$             & 2.32 & 0.83 \\
$\sigma_8$          &$5.2 \times 10^{-2} \, (7.76)$  & 1.15 & 0.96 \\
$H_0$               &$2.4                \, (3.49)$  & 1.23 & 0.94 \\
\hline
\end{tabular}
}
\caption{Estimated accuracy for a $\Lambda$CDM + tensor + massive
  neutrinos model. We assume 3 families of massive neutrinos with
  fully degenerate mass eigenstates, and parametrise the massive
  neutrino contribution to the total energy density by $f_\nu$. For
  $f_\nu$ we quote the $95\%$ upper confidence limit. In parenthesis we
  report the ratio between estimates for this model and estimates for
  the base $\Lambda$CDM +$r$ model for the same channels combination. As
  usual, results of columns 3 and 4 are quoted in units of
  uncertainties for the $70 - 143$ GHz channels.}
\label{tab:numass}
\end{table}

Fitting for the relativistic degrees of freedom and Helium abundance
significantly degrades the accuracy on most parameters, in particular
on $\omega_{\rm dm}$, $\theta$ and $n_{\rm s}$, as shown by the
results of table~\ref{tab:neff70}. In agreement with \citet{hamann:08}
and \citet{ichikawa:08}, we find that Planck alone will be able to
constrain $N_{\rm eff}$ to within $\sim 10\%$, even when
simultaneously determining $Y_{\rm He}$.

\begin{table}
\centerline{
\begin{tabular}{lcc}
\hline
                    & 70 - 143                    & 70 - 217   \\
\hline
$\omega_{\rm b}$          &$2.5 \times 10^{-4} \,  (1.60)$  &  0.93 \\
$\omega_{\rm dm}$    &$4.1 \times 10^{-3} \,  (2.93)$  & 0.87  \\
$\theta  $          &$1.1 \times 10^{-3} \, (3.67)$  &  0.85 \\
$\tau    $          &$4.9 \times 10^{-3} \, (1.04)$  &  0.94 \\
$n_{\rm s}     $          &$8.9 \times 10^{-3} \, (2.23)$  &  0.92\\
${\cal A}_{\rm s}$        & $1.3 \times 10^{-2} \, (1.38)$  &  0.90\\
$Y_{\rm He}        $ & $1.7 \times 10^{-2}$         &  0.88 \\
$N_{\rm eff}       $ & $2.8 \times 10^{-1}$         &  0.87\\
$r$                 & $< 0.029           \, (0.99)$  &  0.86 \\
$\sigma_8$          & $1.1 \times 10^{-2} \, (1.64)$  &  0.88 \\ 
$H_0$               & $1.9              \, (2.75)$  &  0.89\\
\hline
\end{tabular}
}
\caption{As table~\ref{tab:base70}, $\Lambda$CDM + tensor + $N_{\rm
    eff}$ + $Y_{\rm He}$.  Assuming cleaning of the 217GHz channel
    would increase accuracy on most parameters by $\sim 10 \%$. The
    third column reports the errors in units of the uncertainty for
    the base $\Lambda$CDM +$r$ model, see table~\ref{tab:base70}.}
\label{tab:neff70}
\end{table}

\subsubsection{Running spectral index} 

A power law is the simplest possible parametrisation for the power
spectra of both scalar and tensor perturbations. However, several
inflationary theories predicts small deviations from a pure power law,
typically expressed as a logarithmic running of the spectral index,
$n_{\rm run} = {\rm d}n_{\rm s}/{\rm d}log(k)$, at a suitable reference scale, 
$k_{\rm piv}$. Here we choose $k_{\rm piv} = 0.05 {\rm Mpc}^{-1}$. 

\begin{table}
\centerline{
\begin{tabular}{lcccc}
\hline
                      & 70 - 143 & 70 - 100 & 70 - 217   \\
\hline
$\omega_{\rm b}$          &$1.7 \times 10^{-4} $  & 1.42  & 0.88 \\
$\omega_{\rm dm}$    &$1.4 \times 10^{-3} $  & 1.39  & 0.92 \\
$\theta  $          &$3.0 \times 10^{-3} $  & 1.67  & 0.87 \\
$\tau    $          &$5.0 \times 10^{-3} $  & 1.26  & 0.95    \\
$n_{\rm s}     $          &$4.0 \times 10^{-3}  $ & 1.59  & 0.89    \\
$n_{\rm run}$        &$5.8 \times 10^{-3}  $ & 1.33  & 0.91    \\
${\cal A}_{\rm s}$        &$1.1 \times 10^{-2}  $ & 1.25 &  0.95\\
$r$                 & $< 0.037           $ & 2.20  &  0.83 \\
$\sigma_8$          & $6.7 \times 10^{-3} $ & 1.30  &  0.93 \\ 
$H_0$               & $6.9 \times 10^{-1} $ & 1.44 &  0.90\\
\hline
\end{tabular}
}
\caption{Error forecasts for the $\Lambda$CDM + tensor + running
  model.}
\label{tab:run}
\end{table}

\begin{figure*}
\centerline{
\includegraphics[width=6.cm]{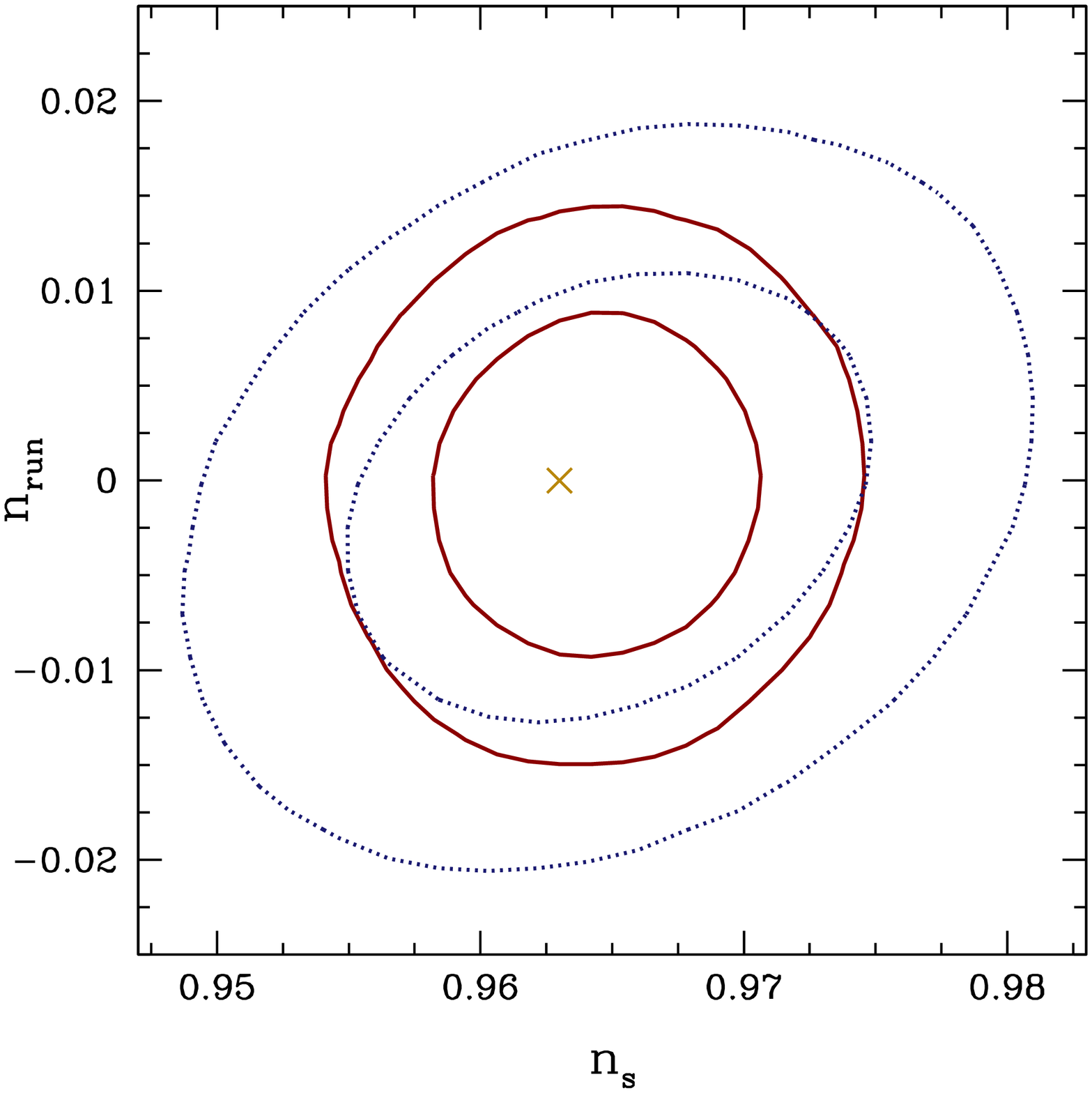}
\includegraphics[width=6.cm]{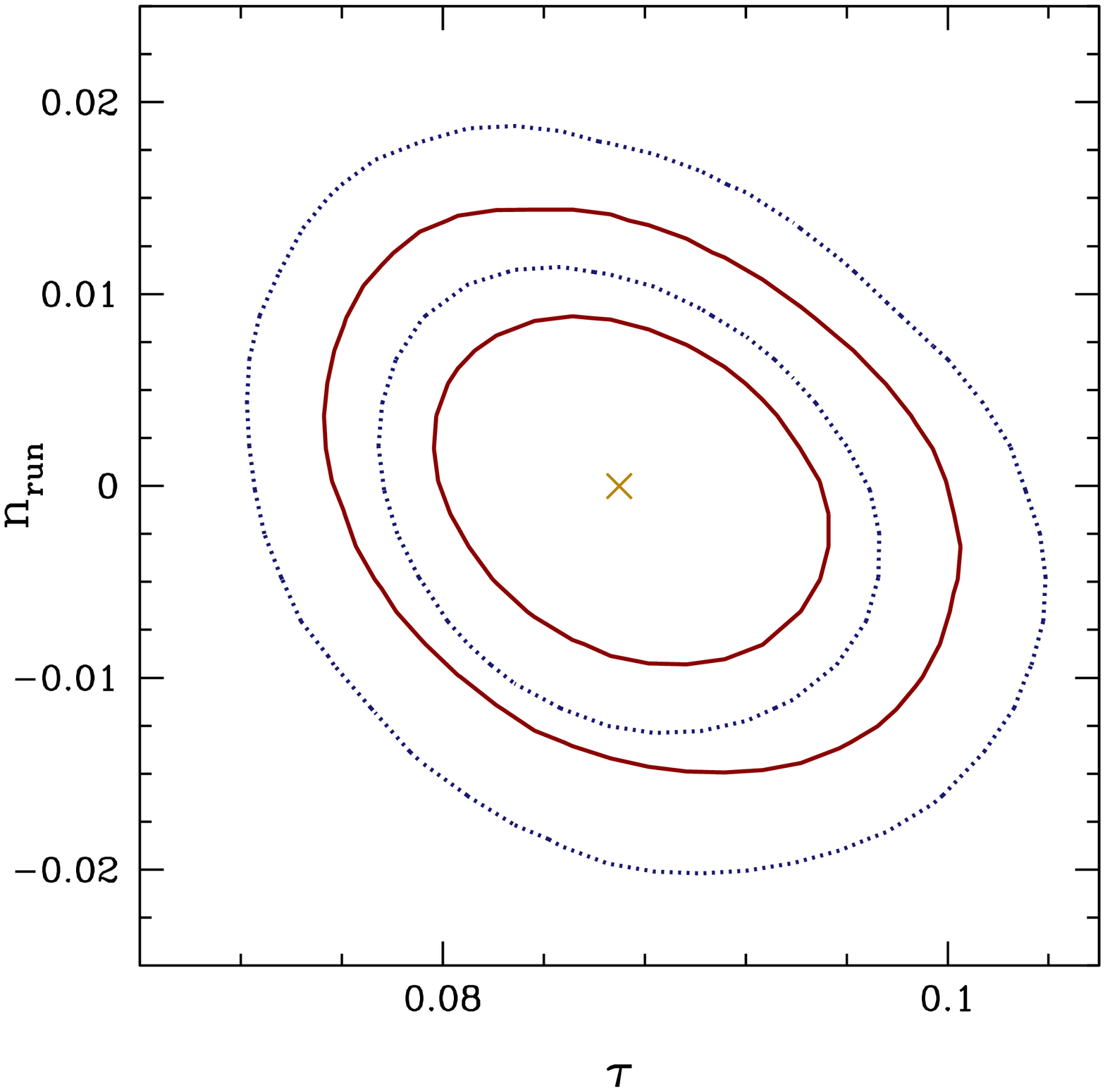}
}
\caption{{\it Left:} 1- and 2-$\sigma$ confidence regions in the $n_{\rm s} -
  n_{\rm run}$ plane, using $70 - 100$ Ghz channels (dashed lines) or
  $70 - 143$ GHz (solid lines) channels.  Using just the $70$ and
  $100$ GHz channels allows for a mild degeneracy between the two
  parameters, which is broken if it is possible to clean the $143$ GHz
  channel. {\it Right:} Confidence regions in the $\tau - n_{\rm run}$
  plane.  The contribution of the $143$ GHz frequency does not help in
  breaking the mild degeneracy between the two parameters.}
\label{fi:ns_nrun}
\end{figure*}

If only the $70$ and $100$ GHz Planck channels are available for
cosmological parameter estimation, we find a moderate degeneracy
between $n_{\rm s}$ and $n_{\rm run}$, similar to the one observed in actual
WMAP5 data. This degeneracy results in an increase of the uncertainty
on $n_{\rm s}$ by $\sim 10 - 15 \%$ over what is obtained for our minimal
parameter set. Including the $143$ GHz channels substantially improves
the determination of the high multipoles, which allows Planck to completely
break this degeneracy, as shown in figure~\ref{fi:ns_nrun}. Even with
the inclusion of the $143$ GHz channel, a minor degeneracy between $n_{\rm
  run}$ and $\tau$ remains, which partly propagates to ${\cal A}_{\rm s}$
and $r$. This degeneracy however, only slightly affects the
constraints on these parameters. The remaining parameters are not
significantly affected by the inclusion of $n_{\rm run}$ in the
parameter set, regardless of the combination of frequencies
considered.

\subsubsection{Tensor spectral index}

One of the most ambitious goals of Planck and other future CMB mission
is the detection of primordial gravitational waves. In the minimal
cosmological model considered here, with the tensor spectral index
fixed to the input value $n_{\rm T} = 0$, $r$ is not significantly
degenerate with the other cosmological parameters. However, even if
the tensor--to--scalar ratio is quite high $r \sim 0.15-0.20$, Planck
will be able to constrain at most the first $10-15$ multipoles of the
B mode spectra. It is then interesting to determine whether Planck will
be able to simultaneously constrain $r$ and $n_{\rm T}$.

We consider two fiducial models with $r = 0.01$ and $r = 0.10$, and
$n_{\rm T}$ determined according to the single field inflation
consistency relation $n_{\rm T} = -r/8$. In both cases, we also
consider a non--zero running of the scalar spectral index $n_{\rm run}
= -0.02$.  The parameter set characterising our model is thus $ \{
\omega_{\rm b}, \omega_{\rm c}, \theta, \tau, n_{\rm s}, n_{\rm run},
{\cal A}_{\rm s}, n_{\rm T}, r\}$. Results of the analysis are
summarised in table~\ref{tab:nt}.

\begin{table}
\centerline{
\begin{tabular}{clcc}
\hline
                      & $r = 0.01$  &   $r= 0.10$    \\
\hline
$\omega_{\rm b}$          &$1.7 \times 10^{-4} $ & $1.8 \times 10^{-4} $  \\
$\omega_{\rm dm}$    &$1.4 \times 10^{-3} $ & $1.4 \times 10^{-3} $ \\
$\theta  $          &$3.1 \times 10^{-3} $ & $3.1 \times 10^{-3} $  \\
$\tau    $          &$5.2 \times 10^{-3} $ & $5.2 \times 10^{-3} $ \\
$n_{\rm s}     $          &$4.0 \times 10^{-3} $ & $4.1 \times 10^{-3} $ \\
$n_{\rm run}$        &$5.8 \times 10^{-3} $ & $6.1 \times 10^{-3} $ \\
${\cal A}_{\rm s}$        &$1.1 \times 10^{-2} $ & $1.1 \times 10^{-2} $\\
$r$                 & $ < 0.16            $ & $ < 0.38  $&\\
$n_{\rm T}$               & $ 0.2              $ & $ 0.2              $ \\
$\sigma_8$          & $6.8 \times 10^{-3} $ & $6.9 \times 10^{-3} $ \\ 
$H_0$               & $7.0 \times 10^{-1} $ & $7.0 \times 10^{-1} $\\
\hline
\end{tabular}
}
\caption{Error forecasts for the $\Lambda$CDM + running + tensor
  spectral index model. Second columns shows results for a fiducial
  model with $r = 0.01$, column 3 refer to $ r= 0.10$. In both cases,
  we considered the $70 - 143 $ GHz channels specifications.}
\label{tab:nt}
\end{table}

\begin{figure*}
\centerline{
\includegraphics[width=6.cm]{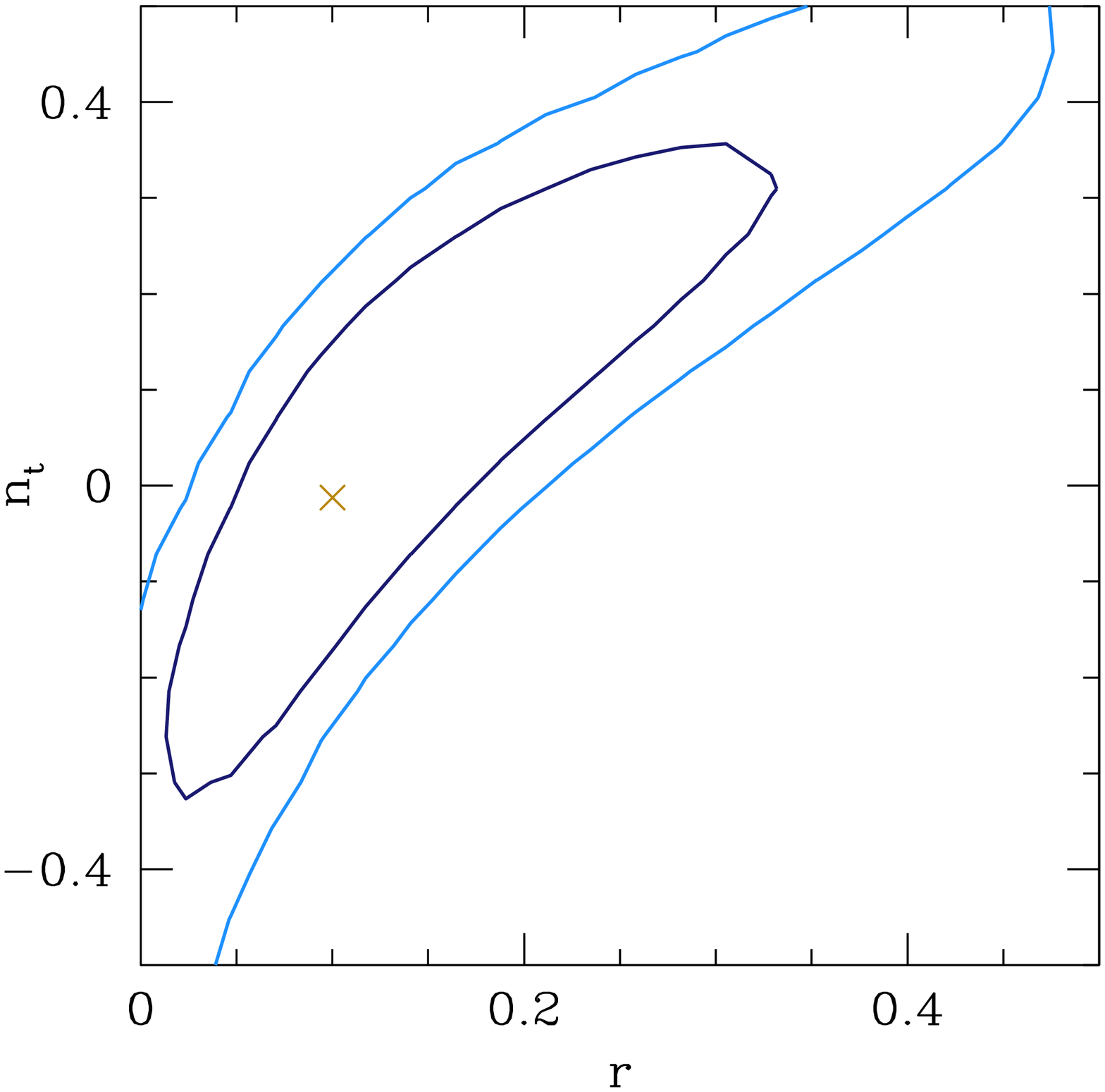}
\includegraphics[width=6.1cm]{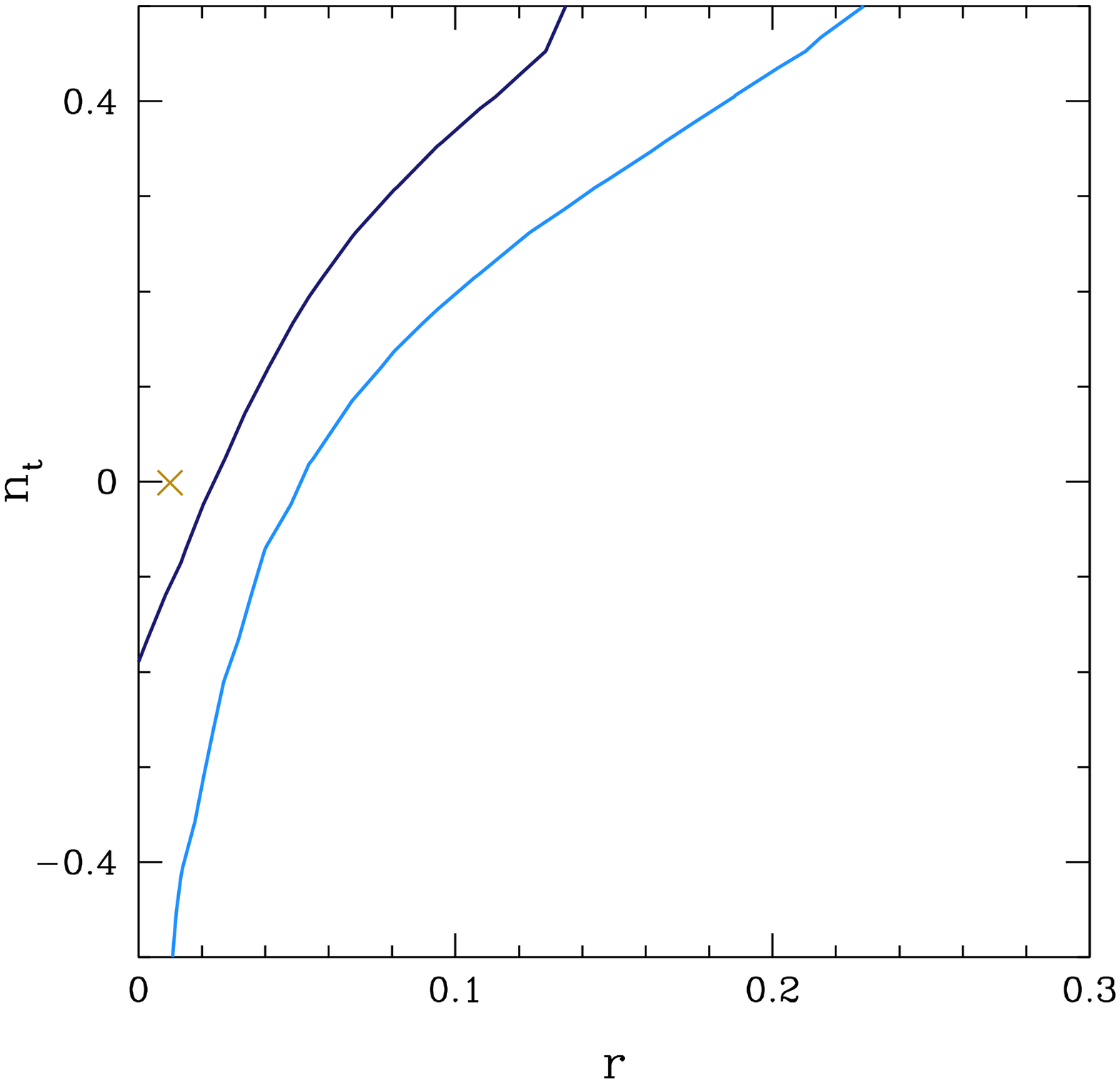}
}
\caption{Degeneracy between the tensor--to--scalar ratio, $r$, and the
tensor spectral index, $n_t$, for fiducial values $\{r = 0.10, n_t =
-0.0125\}$ (left) and $\{r = 0.01, n_t = -0.00125\}$ (right). Even for
moderate values $r \sim 0.10$, Planck does not have sufficient
leverage on the $B$--modes to simultaneously constrain $r$ and
$n_t$.}
\label{fi:nt_vs_r}
\end{figure*}

We find that for Planck the normalisation and spectral index are
completely degenerate, as shown in figure~\ref{fi:nt_vs_r}. Allowing
for $n_{\rm T}$ completely disrupts the instrument's capability for measuring
$r$, even for moderate values of $r = 0.10$, as Planck does not have
sufficient leverage on the $B$--modes to simultaneously constrain 2
tensor mode parameters. Errors on the remaining parameters are similar
to those shown in table~\ref{tab:run}, implying that $n_{\rm T}$ is not
significantly degenerate with the other parameters, in particular with
$n_{\rm s}$ and $n_{\rm run}$.

\section{Planck as support for other missions} 
\label{sec:support}

In the previous section, we have seen that Planck's performance for
wider parameter spaces is often limited by degeneracies inherent to
the CMB data.  Further improvement on parameter constraints will come
not from improved CMB data, but by combination with other experimental
techniques.  In this section, we will use the Fisher matrix formalism
to illustrate constraints for proposed galaxy surveys in combination
with WMAP or Planck.  The Fisher matrix technique has been discussed
in detail by many authors, e.g. \citet{EHT1999}, and we direct the
interested reader to \citet{pritchard2008} for further information on
our Fisher matrix methodology.  Throughout we will assume usage of the
Planck's 70-143 channels for cosmological constraints and consider
WMAP5 and WMAP8 modelled by simply increasing the sensitivity by the
longer integration time.

\subsection{Dark energy}

Dark energy is, perhaps, the most challenging unsolved problem facing
modern cosmology.  However, as we saw in \S\ref{sec:geo}, there is a
strong geometrical degeneracy between $\Omega_\Lambda$, $\Omega_{\rm K}$,
and $w$ present in the CMB.  It is therefore necessary to combine CMB
information with distance measurements in the low redshift Universe in
order to break this degeneracy and constrain dark energy precisely.

A promising technique is the measurement of baryon acoustic
oscillations (BAO) in the galaxy power spectrum.  Since the wavelength
of the BAO is set by the same sound horizon measured in the CMB the
BAO can be used as a ``standard ruler" to measure distances.  CMB
information plays a dual role by calibrating this standard ruler and
providing a distance measurement at high redshifts.

\begin{figure}
\begin{center}
\includegraphics[width=6.cm]{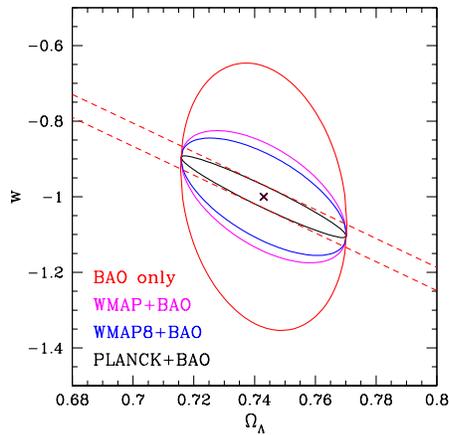}
\caption{$68\%$ confidence ellipses in the $\Omega_\Lambda-w_0$ plane
  for BAO in combination with CMB experiments. Dashed curve indicates
  the constraint from Planck alone with a weak prior on
  $\Omega_\Lambda$.}
\label{fig:wellipse}
\end{center}
\end{figure}

Here we consider constraints from a spectroscopic BAO galaxy survey of
10,000 sq. deg. spanning redshifts $z=0.5-2$. This is a stage IV space
BAO mission in the language of \citet{detf} and a possible
configuration for the proposed JDEM mission.  The $68\%$ confidence
ellipses in the $\Omega_\Lambda-w_0$ plane for this experiment are
shown in Figure \ref{fig:wellipse} and summarised in Table
\ref{tab:dark_energy}, where the constraints have been calculated with
a flat prior and include representative systematic errors on distance
measures to each redshift bin and account for non-linear smoothing of
the acoustic features according to the Fisher matrix prescription
outlined in the Dark Energy Task Force (DETF) report \citep{detf}.  It
is readily seen that the combination of distance measurements at low
and high redshift leads to tight constraints on dark energy
parameters. The low redshift data is essential in fixing the matter
content of the Universe (or equivalently $h$) breaking the geometric
degeneracy.

If we expand our parameter set to allow for evolving dark energy with
equation of state $w(a)=w_0+(1-a)w_a$ \citep{chevallier:01} the
improvement from adding CMB can be quantified by the DETF figure of
merit (FoM), which is defined to be proportional to the inverse area
contained within the $w_0-w_a$ error ellipse.  These are listed in
Table \ref{tab:dark_energy}. WMAP8 provides a $16\%$ increase in DETF
FoM over WMAP5, while Planck is a $\sim2$ fold increase in DETF FoM
over WMAP5.  Thus the goal of making precision measurements of dark
energy very much require precision CMB observations at the Planck
level.
\begin{table}
\caption{Dark energy constraints for BAO + CMB}
\begin{center}
\begin{tabular}{cccc}
Experiment & $\sigma(\Omega_\Lambda)$& $\sigma(w_0)$ & relative FoM \\
\hline
BAO only &0.02 &0.23 & 1\\
BAO + WMAP5 &0.02 &0.12 & 9.6\\
BAO + WMAP8 &0.02 &0.10 & 11.1\\
BAO + Planck &0.02 &0.07 & 20.3\\
\end{tabular}
\end{center}
\label{tab:dark_energy}
\end{table}%

\subsection{Curvature}

Since $\Omega_{\rm K}$ is subject to the same geometric degeneracy as $w$
and $\Omega_\Lambda$ constraining the curvature is vital for obtaining
precision dark energy constraints.  Typically once constraints at the
level of $\Omega_{\rm K}\sim2\times10^{-3}$ are reached this degeneracy is
broken \citep{knox2006}.

\begin{figure}
\begin{center}
\includegraphics[width=6.cm]{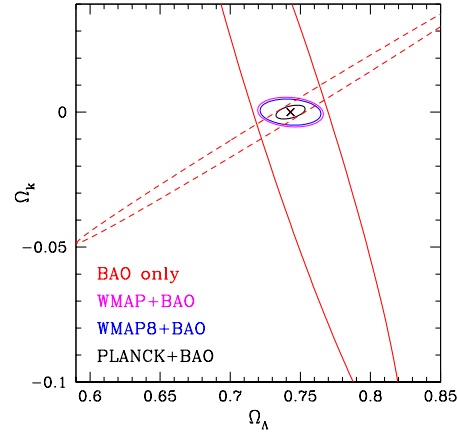}
\caption{$68\%$ confidence ellipses in the $\Omega_\Lambda-\Omega_{\rm K}$
  plane for BAO in combination with CMB experiments. Dashed curve
  indicates the constraint from Planck alone.}
\label{fig:omkellipse}
\end{center}
\end{figure}
We apply the same BAO experiment as in the previous subsection to
calculate the constraints shown in Figure \ref{fig:omkellipse}.  The
strong degeneracy in the CMB data and BAO data have very different
alignments and in combination lead to constraints at the level of
$\Omega_{\rm K}\sim{\rm few}\times10^{-3}$.  Table \ref{tab:curvature}
summarises the results.

Including both a variable dark energy equation of state and curvature
does not degrade these curvature constraints significantly.  Further,
since the curvature is well constrained, we recover constraints on
$\Omega_\Lambda$ and $w$ degraded by $\lesssim 5\%$ from those of
Table \ref{tab:dark_energy} that were calculated with a flat prior.
This illustrates that the degeneracy between curvature and dark energy
parameters is broken by these experiments.

\begin{table}
\caption{Curvature constraints for BAO + CMB}
\begin{center}
\begin{tabular}{ccc}
Experiment & $\sigma(\Omega_\Lambda)$& $\sigma(\Omega_{\rm K})$ \\
\hline
BAO only &0.05 &0.08\\
BAO + WMAP5 &0.015 &0.0037\\
BAO + WMAP8 &0.014 &0.0032 \\
BAO + Planck &0.0066 &0.0017\\
\end{tabular}
\end{center}
\label{tab:curvature}
\end{table}%

\subsection{Inflationary parameters}

Constraining inflationary parameters is of paramount importance in
pointing the way towards a theoretical understanding of first
fractions of a second of the Universe's evolution.  The CMB is
sensitive to the shape of the primordial power spectrum over
wavenumbers ranging from $k=0.002-0.2h\,{\rm Mpc^{-1}}$, since Silk
damping effectively erases information about smaller scales.
Improving constraints on the tilt $n_{\rm s}$ and running $\alpha$ of the
primordial power spectrum will require combining CMB information with
information derived from the matter power spectrum, which extends to
smaller scales increasing the lever arm \citep{adshead2008}.
Precision measurements of $n_{\rm s}$, $n_{\rm run}$, and $r$ will begin to
allow reconstruction of the inflaton potential.

There are currently three techniques proposed for measuring small
scale power: galaxy surveys, the Lyman-alpha forest, and redshifted 21
cm experiments.  The first two techniques are subject to systematic
uncertainties arising from the non-linear evolution of the matter
power spectrum, while the latter is still in its infancy and may be
affected by the imprint of reionisation on the intergalactic medium
(IGM) \citep{fob}.  Perhaps the most promising approach will be high
redshift ($z=2-6$) galaxy surveys, which probe the matter power
spectrum when the scale of non-linearities is small and so may extend
to $k\sim2h\,\iMpc$ or larger.  For this analysis, we consider the
proposed {\em Cosmic Inflation Probe} (CIP), a space based mission to
survey galaxies in H$\alpha$ over the range $z=2-6.5$
\citep{melnick2004}, and assess the importance of CMB support for
placing constraints in the $n_{\rm s}-n_{\rm run}$ plane.  We assume that
CIP will survey 1000 sq. deg. on the sky detecting 200 million
galaxies.  CIP is designed to produce an error ellipse for
inflationary constraints that is comparable to that obtained using
Planck alone.  Since they cover very different scales the combination
is expected to yield precise constraints on inflation.

\begin{figure}
\begin{center}
\includegraphics[width=6.cm]{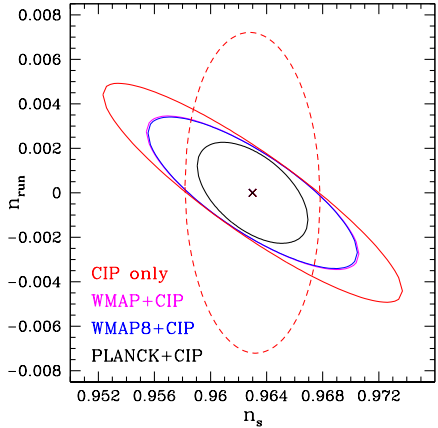}
\caption{Effect of CMB support on CIP for inflation constraints in the
$n_{\rm s}-n_{\rm run}$ plane. Dashed curve indicates the constraint from
Planck alone.}
\label{fig:iellipse}
\end{center}
\end{figure}
Error ellipses for CIP in combination with CMB experiments are shown
in Figure \ref{fig:iellipse}.  Although there is no significant
degeneracy present in the CMB data, for CIP there is a strong
degeneracy between $n_{\rm s}$ and $n_{\rm run}$, since CIP predominantly
measures the power spectrum on scales smaller than the pivot point
$k_{\rm pivot}=0.05\,{\rm Mpc^{-1}}$.  Planck data provides tight
constraints on $n_{\rm s}$ partially breaking this degeneracy and leading to
much improved constraints on $n_{\rm run}$.

\begin{table}
\caption{Inflationary constraints for CIP + CMB}
\begin{center}
\begin{tabular}{cccc}
Experiment & $\sigma(n_{\rm s})$& $\sigma(n_{\rm run})$ & relative FoM \\
\hline
CIP only &0.007 &0.0032 &1\\
CIP + WMAP5 &0.005 &0.0022 &1.47\\
CIP +WMAP8 &0.0049 &0.0022 &1.49\\
CIP +Planck & 0.0026&0.0015 &3.07\\
\end{tabular}
\end{center}
\label{tab:inflation}
\end{table}%
It is apparent that increasing the observation time from 5 to 8 years
for WMAP shows little improvement in the inflationary constraints from
the combination with CIP, Planck is needed.  Constraints are listed in
Table \ref{tab:inflation}.  We may use the generalisation of the DETF
figure of merit to the inflationary parameters \citep{adshead2008} to
quantify this.  For this figure of merit, both WMAP5 and WMAP8 provide
a factor of 1.5 increase over CIP alone, while Planck improves the
inflationary FoM by a factor of 3.0.  The combination of Planck+CIP
measures $n_{\rm run}$ at the level of $10^{-3}$ \citep[in agreement
  with][]{takada:06}, which is roughly the largest value consistent
with simple slow roll inflation models and the current constraints on
$n_{\rm s}$ and is also comparable with that of some more exotic
inflationary models \citep[e.g.][]{large_running,chung2003}.

\subsection{Neutrino masses}

As discussed earlier, the CMB displays a strong degeneracy between
$\omega_{\rm dm}$ and $f_\nu$ since both affect the energy density in
non-relativistic matter before decoupling in a similar manner.  By
adding low red-shift information where neutrinos have become
non-relativistic this degeneracy may be broken \citep[for a review of
neutrino physics, see]{lesgourgues:06}.

In a similar fashion to constraining inflation, constraints on
neutrino mass benefit from constraints on the matter power spectrum on
small scales.  In the relatively near future, surveys such as the {\em
Large Synoptic Survey Telescope}
\footnote{http://www.lsst.org/Science/DETF.shtml} (LSST) are likely to
survey large numbers of galaxy at $z<3$ with photometric redshifts.
These are expected to greatly improve the constraints on neutrino
masses. A similar analysis taking into account the combination of CIP
and Planck was performed by \cite{takada:06}. Neutrinos show
themselves in the matter power spectrum as a suppression of power on
small scales where relativistic neutrinos are able to freestream out
of gravitational potentials smoothing the distribution of matter.  If
neutrinos have mass then at late times they become non-relativistic
and begin to cluster in the gravitational potentials set by the
dominant dark matter component.  Since neutrinos of different masses
would become non-relativistic at different times the amplitude of
power on small scales as a function of redshift is a useful probe of
neutrino mass.

\begin{figure}
\begin{center}
\includegraphics[width=6.cm]{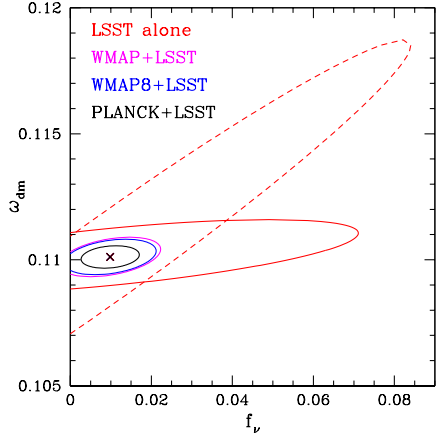}
\caption{Effect of CMB support on LSST for neutrino mass constraints
in the $\omega_{dm}-f_\nu$ plane. Dashed curve indicates the
constraint from Planck alone.}
\label{fig:nuellipse}
\end{center}
\end{figure}
Figure \ref{fig:nuellipse} shows constraints in the
$\omega_{\rm dm}-f_\nu$ plane for $f_\nu=0.01$, corresponding to a total
neutrino mass of $M_\nu=0.1\,{\rm eV}$.  We assume that LSST measures
galaxies with photometric redshifts (with uncertainty $\sigma_z=0.04$)
in six redshift bins over the range $z=0.5-3$. We model the observed
density of galaxies as $n_g(z)=640 z^2 e^{-z/0.35}\,{\rm
  arcmin^{-2}}$.  Note that the uncertainty is to some extent driven
by the photometric redshifts.  If LSST achieved $\sigma_z=0.02$ then
the constraints would be significantly improved and most of the
constraining power would come from LSST alone.

LSST alone does a reasonable job of measuring $f_\nu$, but by placing
a tight constraint on $\omega_{\rm dm}$ it also breaks the degeneracy
present in the CMB data leading to a much tighter constraint on
$f_\nu$.  Once Planck is added to LSST a constraint on $M_\nu$ at the
level of $0.05\eV$ is achievable - sufficient for a detection and
comparable to the value expected from the observed neutrino mass
splittings \citep{fogli2008}.

\begin{table}
\caption{Neutrino mass constraints for LSST + CMB}
\begin{center}
\begin{tabular}{ccc}
Experiment & $\sigma(\omega_{dm})$& $\sigma(f_\nu)$ \\
\hline
LSST only & 0.00097&0.04\\
LSST + WMAP5 &0.00052 &0.0082\\
LSST +WMAP8 & 0.00046&0.0075\\
LSST +Planck &0.00029 &0.0047\\
\end{tabular}
\end{center}
\label{tab:neutrino}
\end{table}%

\section{Conclusions}
\label{sec:concl}

In this paper we performed a MCMC estimate of cosmological
constraints expected from the upcoming Planck data. The actual
contribution by Planck to cosmology will ultimately depend on a number
of factors, especially on systematics control and foreground
cleaning. In this work we adopted a simple approach to foreground
cleaning. We supposed that a number of frequencies will be used to
fully remove foregrounds from the remaining channel, which are then
used to constrain the cosmological parameters, and discussed how such
constraints depend on the frequencies used.

As a general result, we find that a combination of 70, 100, and 143
GHz channels allows Planck to achieve $85 - 90 \%$ of its potential,
corresponding to a factor 3 - 4 improvement over WMAP 5 year results
for the minimal $\Lambda$CDM model. Adding the remaining LFI
frequencies as well as the 217 GHz channel essentially accounts for
all of Planck capabilities in terms of cosmological parameters. Higher
frequencies, while useful for foregrounds control and additional
science, do not significantly increase the accuracy on the main
parameters.

For the cosmology currently preferred by WMAP5 data, Planck will
provide a cosmic variance limited determination of the temperature
power spectrum up to $\ell \simeq 1000$, while the signal--to--noise
ratio will be unity around $\ell \simeq1600 - 1800$ depending on the
channels considered. This high--$\ell$ temperature information is
essential for determination of $n_{\rm s}$ and $\Omega_{\rm b}$
(30-40\% improvement wrt constraints obtaining including only
multipoles up to $\ell_{max}=800$). Multipoles in the range $2000 \la
\ell \la 2500$ only provide a few percent contribution to the accuracy
on cosmological parameters, however this estimate does not consider
gravitational lensing and effects like beam systematics which
introduce correlation between different multipoles. The impact of
multipoles in this range over the final Planck constraints will
strongly depend on the relevance of these effects. Future experiments,
achieving a cosmic variance limited measurement of the TT power
spectrum up to $\ell \simeq 2500$ will improve Planck constraints on
$\omega_{\rm b}$, $n_{\rm s}$ and $\theta$ by $\sim 30 - 40 \%$. We
find that having a beam slightly bigger than that quoted degrades
constraints by $\sim10\%$ on $n_{\rm s}$ and $\Omega_{\rm b}$.
However, adopting a slightly wrong value for the FWHM in the data
analysis does bias the results for several parameters significantly.

Planck will provide a factor of $\sim9$ improvement over WMAP in
determining the upper limit for the tensor--scalar--ratio. Losing the
polarization information from the 143 GHz channel would degrade this
by a factor of two. Being unable to constrain $B$--modes, as results
of foreground contamination would also degrade the error on $r$ by a
factor of $\sim 2$, with a slight dependence on the fiducial value of
the tensor--to--scalar ratio. Planck alone will be unable to constrain
the tensor spectral index $n_{\rm T}$.  Further, if the full 70--143
GHz range is available the degeneracy between $n_{\rm s}$ and $n_{\rm
  run}$ is broken allowing constraints on both a factor of five beyond
those achieved by WMAP. Taken together, Planck will greatly improve
our constraints on the inflationary parameter space.

Significant degeneracies remain in the Planck data due to intrinsic
degeneracies in CMB.  The strong geometric degeneracy in the angular
diameter distance determination remains, although errors on parameters
like $\Omega_{\rm K}$ and $w$ will improve by a factor 2 compared to
WMAP5 thanks to a better determination of the angular scale of the CMB
acoustic peaks. Neutrino properties are much better constrained
(especially $N_{\rm eff}$ for which the current degeneracy is totally
removed), but a strong degeneracy between $f_\nu$ and $\omega_{\rm
dm}$ remains and considering neutrinos in the analysis increases the
estimated error on $\sigma_8$ by a factor $\sim8$.

We have verified that all of these intrinsic degeneracies can be
broken by the addition of extra low-redshift data sets.  Large scale
structure measurements provide a measurement of $\Omega_\Lambda$
breaking the geometric degeneracy between $\Omega_\Lambda$, $w$, and
$\Omega_{\rm K}$ present in CMB data alone. For a JDEM type mission,
Planck improves the constraint on $w$ and $\Omega_{\rm K}$ by almost a
factor of two over using WMAP5 data. Although Planck does not see a
strong degeneracy between $n_{\rm s}$ and $n_{\rm run}$, galaxy surveys like
CIP, which probe smaller scales, do.  Combining CIP with Planck breaks
this degeneracy leading to constraints on the running that are a
factor of 2 better than from Planck alone.  Finally, by measuring the
non-relativistic matter content today, galaxy surveys such as LSST
break the degeneracy between $\omega_m$ and $f_\nu$ improving
constraints on $f_\nu$ by an order of magnitude to the level of the
observed mass splittings.  On all of these parameters using Planck
offers at least a $50-100\%$ improvement over WMAP5, illustrating its
importance as a support for other data sets.

\section*{Acknowledgments}

EP is and NSF--ADVANCE fellow (AST--0649899) also supported by JPL
SURP award 1314616. LPLC and EP were supported by NASA grant
NNX07AH59G and Planck subcontract 1290790 for this work, and would
like to thank Caltech for hospitality during this period. JRP is
supported by NASA through Hubble Fellowship grant HST-HF-01211.01-A
awarded by the Space Telescope Science Institute, which is operated by
the Association of Universities for Research in Astronomy, Inc., for
NASA, under contract NAS 5-26555. We acknowledge the use of the
CosmoMC, PICO and HEALPix packages. The authors thanks Kevin
Huffenberger and Brendan Crill for stimulating discussion on beams
effect.

\clearpage


\end{document}